\documentclass[acmlarge,screen]{acmart}

\AtBeginDocument{%
  \providecommand\BibTeX{{%
    \normalfont B\kern-0.5em{\scshape i\kern-0.25em b}\kern-0.8em\TeX}}}

\usepackage{color}
\usepackage{multirow}
\usepackage{makecell}
\usepackage{subfig}
\usepackage{colortbl}
\usepackage{xcolor}
\usepackage{pgfplots}
\usepackage{pgfplotstable}
\usepackage{tabularx}
\usepackage{longtable}

\pgfplotsset{compat=1.18}

\newcommand{\getcolor}[3]{%
    \pgfmathsetmacro{\colorValue}{100.0*(#1-#2)/(#3-#2)}%
    \edef\temp{\noexpand\cellcolor{blue!\colorValue}\textcolor{black}{#1}}%
    \temp
}

\author{Shuning Zhang}
\email{Zhang.sn314@gmail.com}
\affiliation{
    \institution{Tsinghua University}
    \city{Beijing}
    \country{China}
}

\author{Xin Yi}
\email{yixin@tsinghua.edu.cn}
\affiliation{
    \institution{Tsinghua University}
    \city{Beijing}
    \country{China}
}
\affiliation{
    \institution{Zhongguancun Laboratory}
    \city{Beijing}
    \country{China}
}

\author{Haobin Xing}
\affiliation{
    \institution{Tsinghua University}
    \city{Beijing}
    \country{China}
}

\author{Lyumanshan Ye}
\affiliation{
    \institution{Shanghai Jiaotong University}
    \city{Shanghai}
    \country{China}
}

\author{Yongquan Hu}
\affiliation{
    \institution{University of New South Wales}
    \city{Sydney}
    \country{Australia}
}

\author{Hewu Li}
\affiliation{
    \institution{Tsinghua University}
    \city{Beijing}
    \country{China}
}

\begin{document}

\title[Adanonymizer]{Adanonymizer: Interactively Navigating and Balancing the Duality of Privacy and Output Performance in Human-LLM Interaction}




\begin{abstract}
Current Large Language Models (LLMs) cannot support users to precisely balance privacy protection and output performance during individual consultations. We introduce Adanonymizer, an anonymization plug-in that allows users to control this balance by navigating a trade-off curve. A survey (N=221) measured users' perceived risks and importance of the information for the tasks on nuanced information types. The study also demonstrated that privacy risks were not significantly correlated with model output performance, highlighting the potential to navigate this trade-off. Adanonymizer normalizes privacy and utility ratings by type and automates the pseudonymization of sensitive terms based on user preferences, significantly reducing user effort. Its 2D color palette interface visualizes the privacy-utility trade-off, allowing users to adjust the balance by manipulating a point. The privacy-utility mapping and visualization were set through meticulous iterations and testings. An evaluation (N=36) compared Adanonymizer with ablation methods and differential privacy techniques, where Adanonymizer significantly reduced modification time, achieved better perceived model performance and overall user preference.
\end{abstract}



\begin{teaserfigure}
    \centering
    \includegraphics[width=0.85\textwidth]{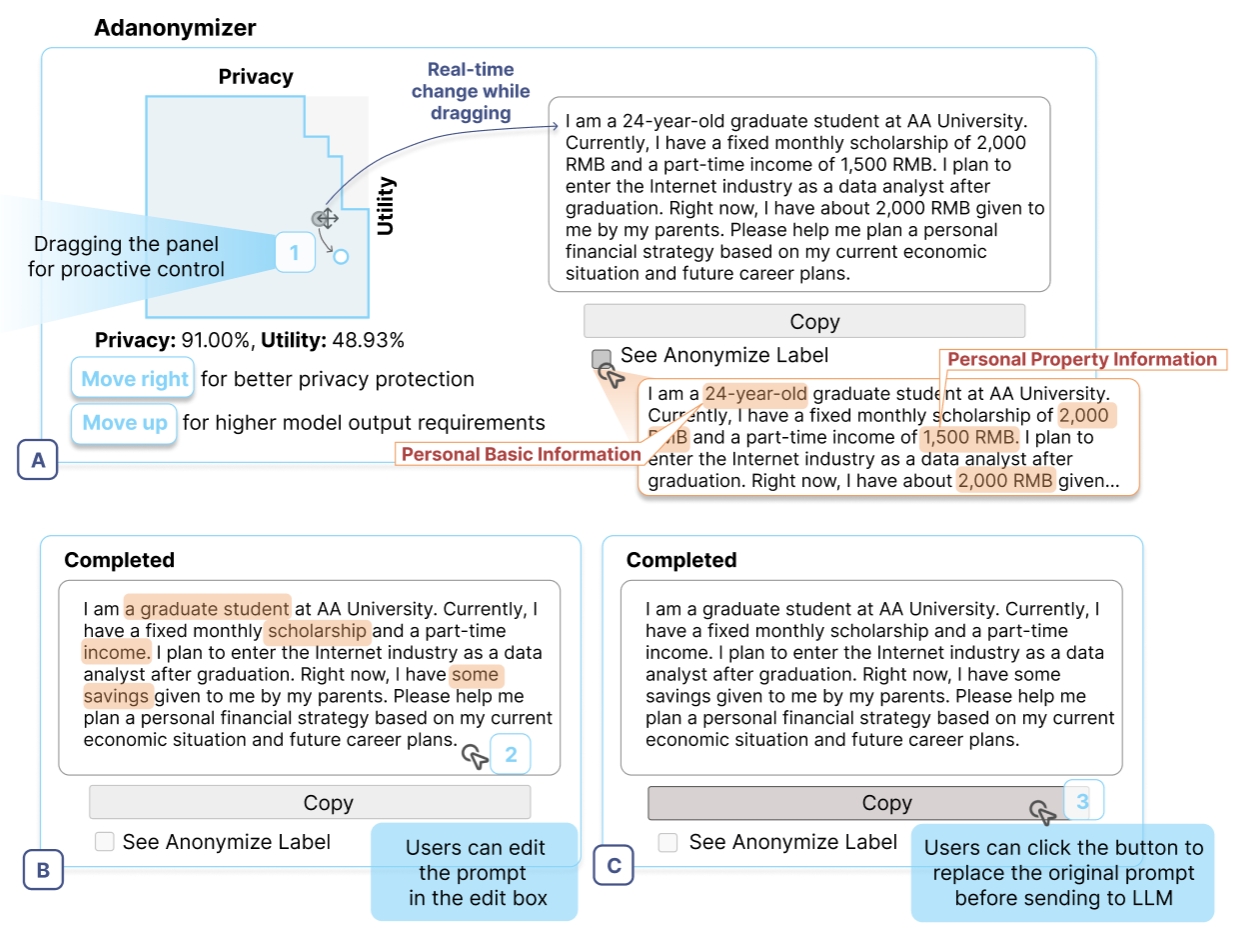}
    \caption{The illustration of Adanonymizer. (A) Users drag the point [1] inside the blue region of the 2D panel to proactively control the privacy protection and model output performance. The output would be real-time re-freshed on the right interface. Users could click the ``See Anonymize Label'' to see the category of the information. (B) Users can view the changed input, their respective information categories and [2] edit them if un-satisfied with the results. (C) Users can click the button to replace the original input.}
    \label{fig:teaser}
\end{teaserfigure}
\maketitle

\section{Introduction}
Large Language Models (LLMs) revolutionized the human-AI interaction paradigm \cite{gao2024taxonomy}, significantly enhancing the interaction experience when users interact with conversational assistants (CAs) driven by LLMs. Users adopted natural language to interact with CAs, facilitating more convenient interaction while in the meantime sacrificing their privacy. 

The potential for information disclosure during user interactions is more pronounced with LLMs compared to traditional text input methods, as highlighted in previous studies~\cite{zhang2024s}. However, despite the rapid adoption of LLMs and the associated rise in privacy risks, systematic investigations into privacy regulations for user inputs remain insufficient. Traditional anonymization techniques like paraphrasing~\cite{staufer2024silencing} fail to adequately account for model performance. Differential privacy approaches~\cite{fernandes2019generalised}, while effective, lack transparency and user control. Furthermore, the sensitivity of text information in differential privacy methods is often task-dependent and opaque~\cite{franzen2022private}. Encryption-based methods~\cite{lin2024promptcrypt} suffer from high computational costs and accuracy trade-offs. The relationship between privacy protection and model performance has long been characterized as a ``trade-off''~\cite{frikha2024incognitext,staab2024large}, where stringent privacy requirements often degrade model effectiveness. The trade-off has significantly hindered the widespread adoption of privacy-preserving techniques for text-based models~\cite{lison2021anonymisation}.

Collaborative interaction allows users and LLMs to jointly modify content, which offers poettnail to solve the above challenge as users are the most important judger of utility and privacy protection. Enabling users to better manage privacy protection and model output performance could better fit for their personalized preferences~\cite{asthana2024know,zhang2024s}. However, how users could effectively balance privacy protection and model output during text-based human-LLM interaction was less explored. Hence this work addresses the key challenge: How can privacy protection and model output performance be balanced through collaborative interaction between humans and AI in human-LLM Interactions? Specifically, we proposed three research questions:

\begin{itemize}
    \item RQ1: How to include users' privacy-utility trade-off cognition into the system design in a quantitative manner?
    \item RQ2: How to design a collaborative interaction interface that effectively and efficiently balances both?
    \item RQ3: How would users balance privacy and utility through using this sytem in human-LLM interaction?
\end{itemize}  

To investigate RQ1, we conducted a questionnaire study (N=221) to examine users' privacy perceptions regarding the input they provide during human-LLM interactions. We designed a set of vignettes on Personal Information (PI) disclosure, framed with specific scenarios, as part of a questionnaire to assess participants' perceived privacy risks, utility and the frequency with which various types of PI were included in their inputs. Privacy risks were evaluated based on two related concerns: information leakage and retaining risks. The results revealed participants' ``Privacy Paradox''~\cite{nissenbaum2011contextual}, who input specific types of private data frequently despite recognizing its privacy implications. Users were primarily motivated by convenience, often overlooking the associated privacy risks~\cite{zhang2024s}. We also found no significant correlation between privacy risks and perceived model output performance degradation when private information was excluded. Personal health and identification data presented higher privacy risks and had a notable impact on model performance, while basic personal information posed lower risks and had minimal effect on output quality. This divergence between privacy concerns and utility informed the design of Adanonymizer.

We introduce Adanonymizer, a collaborative anonymization technique with a color palette-like design, allowing users to balance privacy and output performance (see Figure~\ref{fig:teaser}). Following the ``notice and choice'' principle~\cite{nissenbaum2011contextual}, Adanonymizer visualizes the privacy-utility trade-off as a curve on a two-dimensional plane. The visualization adopts normalized linear mapping for clearer communication and selects continuous panel for smoother control. It allows users to intuitively explore the trade-off by selecting points below or near the curve and receiving immediate feedback, similar to adjusting a color palette. Users can click near the curve to find candidates that balance privacy and model output performance more effectively. Adanonymizer anonymizes different types of information based on users' privacy and performance preferences, as well as the results from Study 1, with the option for further modifications if necessary. Adanonymizer identifies personal information through a prompt-based approach on local quantized models and selectively anonymizes it according to the chosen proportion. 

A usability evaluation study with 36 users compared Adanonymizer with its ablation techniques and a Differential Privacy (DP)-based baseline technique. We selected the personal consultation task under three different contexts \cite{zhang2024s} (work-related, academic-related and life-related) where inputs often contain private personal information. Adanonymizer outperformed other baseline techniques in terms of modification time, perceived output performance and perceived satisfaction. Additionally, users favored Adanonymizer for its enhanced experience, perceived control and transparency.

To sum up, the contributions of this work were three-folded:
\begin{itemize}
\item We conducted a survey (N=221) on users' privacy risks in LLM interactions, quantifying the privacy paradox and examining the trade-offs between information sensitivity, usage frequency and importance.
\item We introduced Adanonymizer, employing a novel 2D color palette metaphor to help users balance privacy and performance with immediate feedback, offering new insights into privacy-utility trade-offs.
\item We tested Adanonymizer in real scenarios (N=36), revealing efficient privacy-utility balancing strategies, categorized into three distinct exploration patters, with implications for future privacy-utility design.
\end{itemize}
 
\section{Related Work}

\subsection{Privacy Risk in LLM-based Conversational Assistants}\label{sec:rw_privacy_risk_llm}

LLM-based CAs face privacy risks both during training and inference, with the paper focusing primarily on the latter. During training, these agents rely on LLMs, which are trained using vast amounts of data, including user interactions~\cite{pahune2023several}. While this data-driven approach enhances the agents' conversational abilities, it introduces LLMs' memorizing risk which could potentially disclose personally identifiable information (PII) from the training data~\cite{brown2022does,peris2023privacy}. Despite safety measures, LLMs, such as ChatGPT may inadvertently reveal sensitive information when prompted with carefully constructed querie~\cite{kim2024propile}.

During inference, as users also shared sensitive data~\cite{zhang2024s}, LLMs may collect these data for target advertising, commercial use or used these data for training. Beyond the LLM-specific training use, other privacy risks within AI literature also threatens user-LLM interaction~\cite{lee2024deepfakes}, including invasion, data collection, data dissemination and data processing risks~\cite{solove2005taxonomy,otrel2010taking,ping2018automatic,milmo2021amazon}. Invasion risks often involve the disruption of personal solitude, such as through targeted advertising or surveillance enabled by AI~\cite{otrel2010taking,ping2018automatic,milmo2021amazon}. Data collection risks, which pertain to the process of gathering personal data, can exacerbate surveillance concerns~\cite{solove2005taxonomy}. Data processing risks emerge from the use, storage, and manipulation of personal data, including aggregation, where disparate data points are combined to make inferences beyond what is explicitly stated~\cite{solove2005taxonomy}. Finally, data dissemination risks, including exposure, distortion and unauthorized use, can lead to privacy violations, particularly when AI-generated content is misused~\cite{levin2017new,burgess2021biggest,ayyub2018india}. These risks are further amplified when user input is exposed.

\subsection{Privacy Preserving Methods for Human-LLM Interaction}\label{sec:rw_text_anonymization}

During the interaction phase, privacy preserving methods contained anonymization, differential privacy and encryption. Anonymization needs arose with text-based interaction \cite{rubin1993statistical}, which traditionally included anonymization and pesudo-anonymization \cite{lison2021anonymisation}. The conventionally anonymization pipeline first identifies entities and then applies replacement strategies \cite{rubin1993statistical}. However, the anonymization process often failes to consider the utility \cite{lison2021anonymisation}. With advancements in adversarial learning \cite{li2018towards,coavoux2018privacy} and prompt-based rewriting \cite{staab2024large}, researchers seek methods to improve the simple replacement strategies \cite{li2018towards}, such as adversarial learning \cite{li2018towards,coavoux2018privacy} and reinforcement learning \cite{mosallanezhad2019deep}. More advanced approaches merge differential privacy and adversarial learning based methods \cite{phandifferential,alnasser2021privacy}. Unfortunately, these methods heavily relies on the computational resources, making it infeasible for end-users. 

Differential privacy was originally proposed for obfuscating the probability of being identified in a set \cite{dwork2006differential}, which depended on adding noise (often sampled via a Laplace distribution) to the original dataset. Local differential privacy (LDP) was widely adopted in various natural language processing (NLP) tasks \cite{qu2021natural}, where more flexible metrics were consecutively proposed to improve the privacy protection effect \cite{feyisetan2020privacy,fernandes2019generalised,feyisetan2019leveraging}. However, these methods modeled the privacy and utility from a calculative perspective, which heavily relies on the accuracy of modeling. Previous literature however proved estimating the population for differential privacy is hard \cite{mehner2021towards}. Worse still, these calculative methods were in-transparent and thus unsuitable for end-users' perception \cite{nanayakkara2023chances,cummings2021need,gadotti2022pool}. 

In encryption, techniques such as homomorphic encryption \cite{acar2018survey} and secure multiparty computation \cite{goldreich1998secure} methods allow servers to perform private inference on encrypted user queries. Nevertheless, the practicality of this method is limited by its high computation overheads \cite{luo2024secformer}, making it hard to employ for human-LLM interaction. Alternative approaches, like fine-tuning or representation learning to simulate the ``encryption'' effect, also suffer from high computational costs and accuracy loss \cite{zeng2024privacyrestore,chen2023hide}.

Interaction design and interface design also play a crucial role in addressing privacy concerns during human-LLM interactions. Previous research has highlighted these aspects, particularly in tasks involving text rewriting and generation. For example, Lee et al. \cite{10.1145/3613904.3642697} emphasized users' concerns about data handling throughout the writing process, while Lin et al. \cite{10.1145/3613904.3642217} introduced Rambler, a speech-based writing interface powered by LLMs, featuring tools like the Semantic Merge button, New Ramble button, and Semantic Zoom slider. These studies provide the interface design inspirations for this work, which adopted a user-centric interaction perspective in addressing privacy risks.

\subsection{Privacy-utility Trade-off in Privacy Regulation of Text}\label{sec:rw_privacy_utility}

Privacy and utility trade-off has long been identified and debated in the context of text document privacy preservation \cite{mireshghallah2021privacy}. Differential privacy is one most outstanding method for modeling the trade-off through the parameter epsilon \cite{dwork2006differential}. However, the determination and interpretation of the trade-off proved challenging \cite{asikis2020optimization}, and opaque to users \cite{nanayakkara2023chances,franzen2022private}. 

Besides differential privacy, researchers have modeled the privacy-utility trade-off by assigning distinct metrics \cite{wunderlich2022privacy,feyisetan2020privacy} and developing different optimization algorithms \cite{shi2021selective}. Privacy is frequently tied to differential privacy guarantee \cite{adelani2020privacy} or specific attack success rates (ASR) \cite{shen2024fire}, whereas utility is often measured by task performance \cite{staab2024large}, loss-based gradient \cite{shen2024fire}, control of specific attribute \cite{frikha2024incognitext} or even directly estimated through LLMs \cite{frikha2024incognitext,staab2024large}. Methods that calculate utility based on single-task performance \cite{mosallanezhad2019deep} struggle to generalize across multi-task scenarios common in human-LLM interactions. Prior research has demonstrated the feasibility of computational~\cite{shen2024fire} and LLM-based methods~\cite{staab2024large} for calculating privacy and utility. However, as the privacy calculus theory suggests \cite{laufer1977privacy}, ``privacy and utility represent the gain and loss a person experiences regarding the potential privacy risks and the task completion benefits.'' Modeling this trade-off computationally is difficult, and our work explores the human-centric aspects of the privacy-utility balance.

Although the human-centric aspects of privacy-utility trade-off in text anonymization was under-explored, there have been early efforts. Ragan et al. \cite{ragan2018balancing} used anonymization and correlation mark-ups to determine how much information could be anonymized without affecting human decision accuracy in record-linkage tasks. However, their focus was on database information, which differs significantly from natural language interaction. Thus, this paper initiated the first study to investigate the human-in-the-loop control of the privacy-utility trade-off in natural language contexts.

\section{Study 1: Evaluating Privacy Risks and Impact of Private Information Removal on Model Performance in Human-LLM Interaction}\label{sec:study_1}

In this section, we apply Contextual Integrity theory \cite{nissenbaum2004privacy} to examine the sensitivity and perceived frequency of various types of potentially disclosed information.

\subsection{Recruitment and Participants}

We recruited 221 Chinese participants (123 females, 98 males) through distributing the questionnaire on an online recruiting platform \footnote{\url{https://wjx.cn}, last accessed at Sep 12th, 2024}. To ensure diversity, we distributed the questionnaire in different time across three days. Participants were required to be between 18 and 65 years old. 141 participants were 26 to 35 years old with 47 participants 36 to 45 years old. 56 were with a manufacturing occupation including mechanical engineers, factory works, etc. 45 were from information technology occupation such as software development engineers and system administrators. Others are from human resources, retail, service industry or other fields. 171 participants held a bachelor degree, and 32 had a master's degree or higher. The detailed demographics of participants were shown in Table~\ref{tbl:study1_demographics} in the appendix. Each participant who completed the experiment received 20 RMB as compensation according to the local wage standard.

\subsection{Scope of Personal Information}

Personal data, as defined in General Data Protection Regulation (GDPR), refers to ``any information related to an identified or identifiable natural person''. In designing scenarios and personal information (PI) categories, we followed Contextual Integrity~\cite{nissenbaum2004privacy}, China's classification criteria\footnote{\url{https://www.tc260.org.cn/upload/2021-12-31/1640948142376022576.pdf}} and prior studies~\cite{richthammer2014taxonomy,richthammer2013taxonomy,chua2021effects,malgieri2016property}. We synthesized classifications from these sources, noting that varying interpretations of personal and privacy information result in inconsistent categories, often tailored to specific contexts, such as social media applications. To address this, we consolidated information from multiple sources into 14 broader categories, providing a comprehensive classification.

These categories include: personal basic information, identification information, online identity, health and physiological data, medical information, education and employment data (covering both education and employment), financial and property data (including financial accounts, transactions, assets, and loans), records, location data, exercise data, and other personal information. A detailed description of these types is provided in Appendix Section~\ref{appen:information_type}.

\subsection{Experiment Design}
We conducted a one-factor within-subjects study with \emph{PI} as the sole factor, examining 14 PI categories (e.g., including personal identity, exercise information) and 74 specific types (e.g., name, age) as detailed in the previous subsection.

To ensure participants understood privacy risks, we provided descriptions of information leakage and model retraining risks in the questionnaire.\textit{Information Leakage Risk refers to the potential disclosure of sensitive information during usage, such as business data or academic assignments, to malicious third-parties. Retraining Risk arises when users' sensitive input data is used by service providers to improve models through further training}. We presented several examples classified by scenario to illustrate each PI, with each example including multiple PI types. We additionally added test questions before the rating entries to ensure the understanding of participants. We measured privacy risk, perceived usage frequency, and perceived utility degradation using a 7-point Likert scale. In total, there were 222 Likert scale entries (3 factors $\times$ 74 PI types). We manually crafted a few cases for illustration of the perceived utility degradation. Detailed information about the questionnaire can be found in Section~\ref{sec:appendix_questionnaire}.

\begin{itemize}
    \item perceived privacy risk: the privacy risk when PIs from your input is used for training, fine-tuning, or could be leaked by LLMs. This risk considers only the sensitivity of the information, not how often it is shared.
    \item perceived usage frequency: how frequent you disclose the PI in the input of your interaction with LLM-based products.
    \item perceived utility degradation: How would the model performance potentially be influenced without this provided information?
\end{itemize}

\subsection{Pilot Studies}

Due to the complexity of LLM interactions, traditional PI classifications do not fully capture all relevant types. To address this, we conducted three pilot studies with 10 participants each, refining the set of PI types specific to LLM interactions. Participants were asked open-ended questions to suggest any additional sensitive information they might disclose, and were compensated with 1RMB for each type disclosed. We followed data saturation theory~\cite{fusch2015we}, halting when no new disclosures emerged. 

The pilot studies revealed no additional types of information, confirming the comprehensiveness of our set. We also revised privacy risk explanations and data descriptions based on participant feedback. In the final round, participants demonstrated clear understanding of the risks and scenarios, with no further misunderstandings. 


\subsection{Procedure}

Prior to the experiment, participants were briefed on its purpose and risks, and provided informed consent. They could withdraw at any time. The questionnaire contained 222 items, taking an average of 1756 seconds (SD=710) to complete. No participant dropped out. To ensure quality, we included 4 comprehension checks on PI. As a result, 9 of 221 participants (4.07\%) were excluded, leaving 212 valid responses. 

\subsection{Results}

\begin{figure}[!htbp]
    \includegraphics[width=\textwidth]{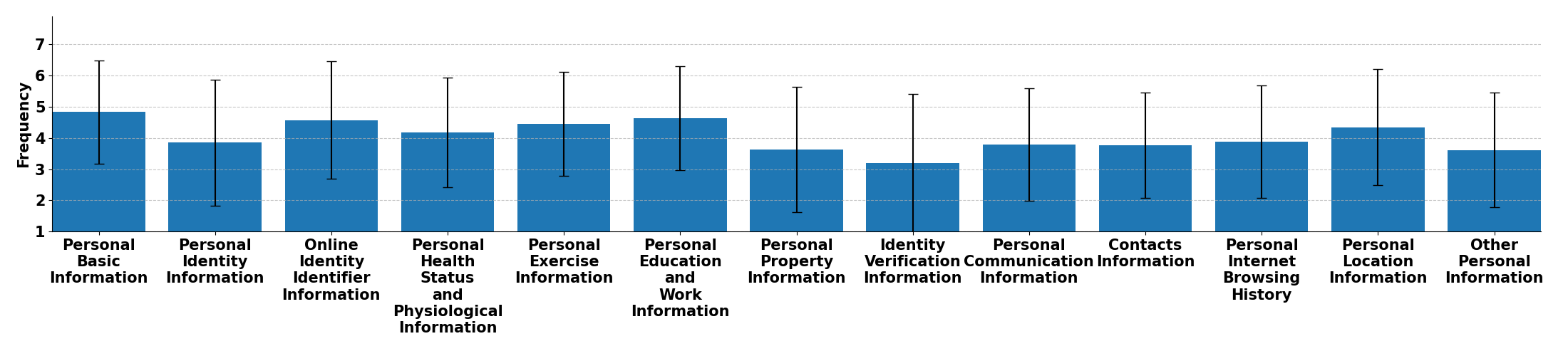}
    \caption{The illustration of private information's appearance frequency, 7: most frequent, 1: least frequent. The errorbar indicated one standard deviation.}
    \label{fig:heatmap}
\end{figure}

This section presents the perceived privacy risk ratings and perceived frequency distributions for different information types, followed by an analysis of their relationship, including the link between privacy risk and perceived utility.

\subsubsection{Perceived Privacy Risk and Usage Frequency of Different Scenarios and PI}\label{sec:study1_privacy_risk}
Table~\ref{tbl:privacy_risk} and Figure~\ref{fig:heatmap} presented the privacy risks and appearance frequencies with different scenarios and PI types separately. Our findings indicate that participants were most concerned about \textit{Identity Verification Information} (M=5.86), while \textit{Personal Exercise Information} (M=3.39) attracted the least attention. The top three perceived high-risk categories—\textit{Identity Verification Information} (M=5.86), \textit{Online Identity Identifier Information} (M=5.74), and \textit{Personal Property Information} (M=5.57)—likely rank highest because their exposure could lead to significant financial or personal loss. Conversely, \textit{Personal Exercise Information} (M=3.39), \textit{Personal Basic Information} (M=4.32), and \textit{Personal Health Status and Physiological Information} (M=4.94) were rated as lower risk, possibly due to their frequent use and the minimal harm caused by their disclosure. In terms of appearance frequency, \textit{Personal Basic Information} (M=4.82) appeared most often, followed by \textit{Online Identity Identifier Information} (M=4.56) and \textit{Personal Education and Work Information} (M=4.63). These differences suggest that users may choose to anonymize information based on their privacy risk tolerance and usage patterns.

\begin{table}[h!]
\centering
\caption{Privacy risk of personal information data, represented by a purple grid, where opacity varies based on information sensitivity. Darker shades indicate higher sensitivity.}
\label{tbl:privacy_risk}
\begin{tabular}{>{\raggedright\arraybackslash}p{7cm} c c}
    \toprule
    \textbf{Category} & \textbf{Mean} & \textbf{Standard Deviation} \\
    \midrule
    Personal Basic Information & \getcolor{4.240}{3.2}{6.5} & 1.797 \\
    Personal Identity Information & \getcolor{5.790}{3.2}{6.5} & 1.514 \\
    Online Identity Identifier Information & \getcolor{5.878}{3.2}{6.5} & 1.141 \\
    Personal Health Status and Physiological Information & \getcolor{4.768}{3.2}{6.5} & 1.634 \\
    Personal Exercise Information & \getcolor{3.327}{3.2}{6.5} & 1.543 \\
    Personal Education and Work Information & \getcolor{5.069}{3.2}{6.5} & 1.428 \\
    Personal Property Information & \getcolor{5.619}{3.2}{6.5} & 1.352 \\
    Identity Verification Information & \getcolor{5.966}{3.2}{6.5} & 1.508 \\
    Personal Communication Information & \getcolor{5.188}{3.2}{6.5} & 1.381 \\
    Contacts Information & \getcolor{5.110}{3.2}{6.5} & 1.373 \\
    Personal Internet Browsing History & \getcolor{5.122}{3.2}{6.5} & 1.486 \\
    Personal Location Information & \getcolor{4.398}{3.2}{6.5} & 1.663 \\
    Other Personal Information & \getcolor{3.946}{3.2}{6.5} & 1.647 \\
    \bottomrule
\end{tabular}
\end{table}

\begin{figure}[!htbp]
    \includegraphics[width=0.85\textwidth]{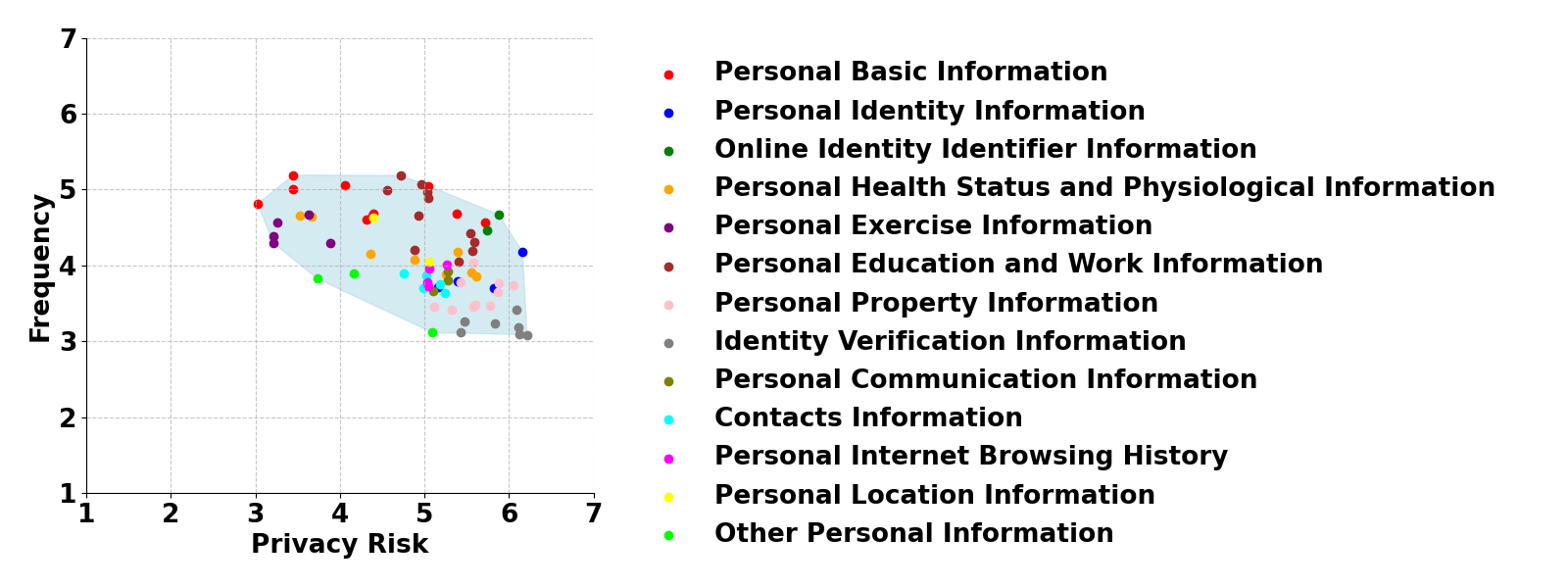}
    \caption{The trade-off between privacy risk (1: least private, 7: most private) and usage frequency (1: least frequent, 7: most frequent). Different information is aggregated into different categories with different colors. We used light blue shading to highlight the envelope, allowing for observation of the overall trend.}
    \label{fig:privacy_freq_tradeoff}
\end{figure}

Figure~\ref{fig:privacy_freq_tradeoff} illustrates the relationship between information sensitivity and appearance frequency. While there is a general negative correlation between these variables ($R^2=-0.56$), several types of information present both high privacy risks and frequent appearances. For instance, \textit{Personal Educational and Work Information} were frequently cited by participants (M=4.63) and associated with significant privacy risks (M=5.11). Similarly, \textit{Only Identity Identifier Information} was mentioned frequently (M=4.56) and linked to high privacy risks (M=5.80). In contrast, \textit{Identity Verification Information}, despite posing a higher privacy risk (M=5.89), appeared less frequently in responses (M=3.20). This suggests that participants did not fully account for the relationship between usage frequency and privacy risk, often disclosing highly sensitive information, reflecting the so-called ``Privacy Paradox''~\cite{nissenbaum2011contextual}.
\subsubsection{The Balance Between Perceived Privacy Risk and Model Performance}\label{sec:study1_tradeoff}



\begin{figure}[!htbp]
    \includegraphics[width=0.85\textwidth]{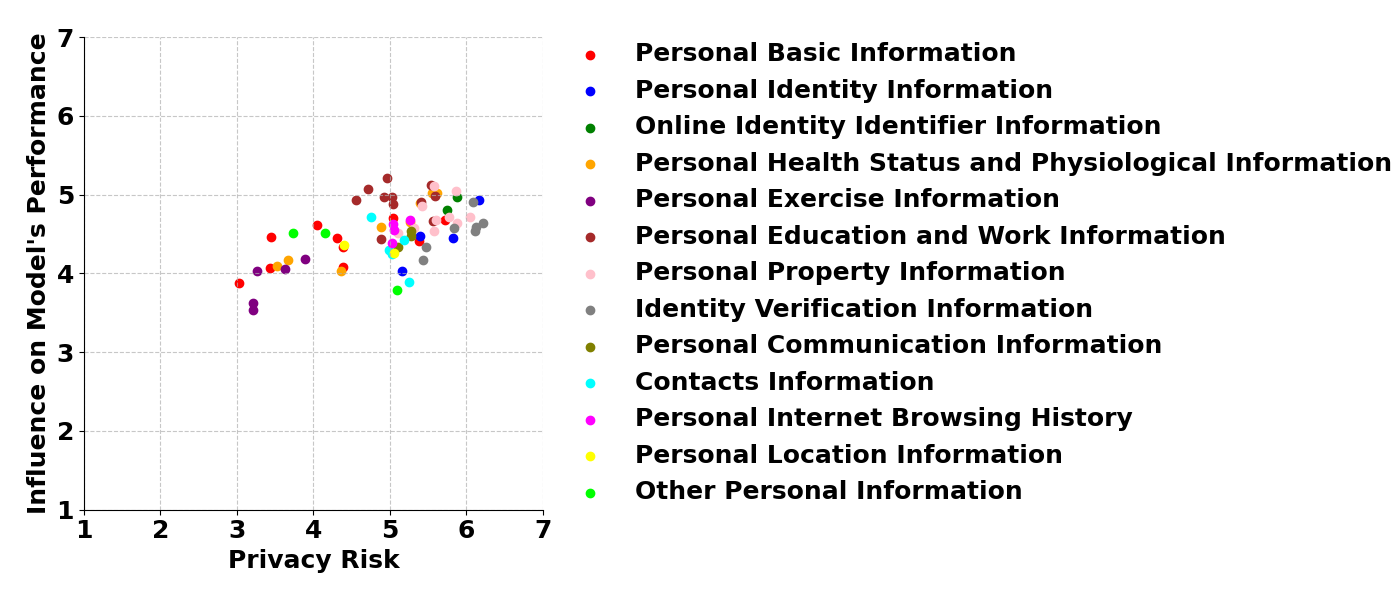}
    \caption{The trade-off between privacy (1: least private, 7: most private) and the influence on model's performance (1: least influential, 7: most influential) for different information. Different information is aggregated into different categories with different colors.}
    \label{fig:privacy_utility_tradeoff}
\end{figure}

\textit{The influence on model performance} was defined by whether the deletion of specific information affected the model’s output. As shown in Figure~\ref{fig:privacy_utility_tradeoff}, \textit{Personal Education and Work Information} (M=4.91), \textit{Online Identity Identifier Information} (M=4.80), and \textit{Personal Property Information} (M=4.74) ranked as the top three categories influencing model performance the most. In contrast, \textit{Personal Exercise Information} (M=3.84), \textit{Personal Location Information} (M=4.25), and \textit{Contacts Information} (M=4.29) were the least influential.

The correlation between information types and model performance showed limited alignment with privacy risk (see Figure~\ref{fig:privacy_utility_tradeoff}). For instance, \textit{Personal Identity Information} had a high privacy risk (M=5.45) but a lower influence on model performance (M=4.31). In contrast, \textit{Personal Health and Physiological Information} had a lower privacy risk (M=4.94) but a higher influence (M=4.61). Other categories, such as \textit{Personal Basic Information} (M=4.32, M=4.36) and \textit{Personal Exercise Information} (M=3.39, M=3.84), were rated low in both privacy risk and influence on the model's performance, whereas \textit{Personal Property Information} had high scores for both privacy risk (M=5.57) and influence (M=4.74). These findings suggest that users could customize their preferences, balancing privacy protection and model performance.


\section{Design of Adanonymizer}

\subsection{Design Goals}
Building on prior work~\cite{lison2021anonymisation} and aiming to balance privacy protection and model performance, we introduced the following design goals:

\begin{itemize}
\item \textbf{DG1: Achieve both high usability and effective privacy protection.} The anonymization technique needed to both guarantee a high usability and in the same time achieve effective privacy protection, as opposed to the past technique which only protect the privacy \cite{lison2021anonymisation} without concerning the usability.

\item \textbf{DG2: Proactive control of participants.} Users were found to favor manual control over their privacy~\cite{wijesekera2017feasibility,reinhardt2021visual} however also exhibited privacy paradox~\cite{nissenbaum2011contextual}, where they seldom manually edit the results. The system should provide users with the control of privacy protection levels but automatically replace the sensitive phrases. This differed from the fully automatic design~\cite{wei2018autoprivacy,ragan2018balancing} and the manual design~\cite{zhou2016enhancing,cranor2006user} through granting users and sense of convenience, control and agency~\cite{zhang2019proactive}. The comparison of different proactive levels were detailed in Section~\ref{sec:study2_results}. 

\item \textbf{DG3: Intuitive and easy to understand.} The anonymization and visualization process should be transparent and easy to understand for participants. In particular, the anonymization and the guarantee of privacy should be comprehensible for participants. 
\end{itemize}

\subsection{Proactiveness of Adanonymizer}


The core of Adanonymizer lies in its human-AI collaborative approach to control privacy, featuring three operational modes that offer varying levels of user involvement. In the most automated mode (denoted as \textit{Automated Mode}), the system autonomously determines which information to anonymize, minimizing user involvement. Conversely, the fully manual mode (denoted as \textit{Manual Mode}) grants users complete control over anonymization decisions. The intermediate mode (denoted as \textit{Single Mode}, which means users could control a single dimension) allows users to make certain decisions—specifically regarding privacy control—while the system manages the rest.

User control is facilitated through a slider, which adjusts the balance between privacy control and system usability. In the \textit{Automated Mode}, the system governs both privacy and usability, leaving users with no control over individual pieces of information. In the \textit{Manual Mode}, users have full autonomy, managing both privacy and usability independently, with sliders arranged either side by side or orthogonally to form a control panel. For the \textit{Single Mode}, users control privacy settings, while the system automatically optimizes usability. The final interface of Adanonymizer was selected as the \textit{Manual Mode} with no proactive control, providing users with a strong sense of control and the ability to balance privacy protection with model performance. Notably, the evaluation study also compared the privacy protection effectiveness and model output performance of different configurations.

\subsection{Palette Design}

\begin{figure}[!htbp]
    \centering
    \subfloat[Discrete.]{
        \includegraphics[page=2,width=0.3\textwidth]{figures/Adanonymizer.pdf}
    }
    \subfloat[Continuous with magnet effect.]{
        \includegraphics[page=1,width=0.3\textwidth]{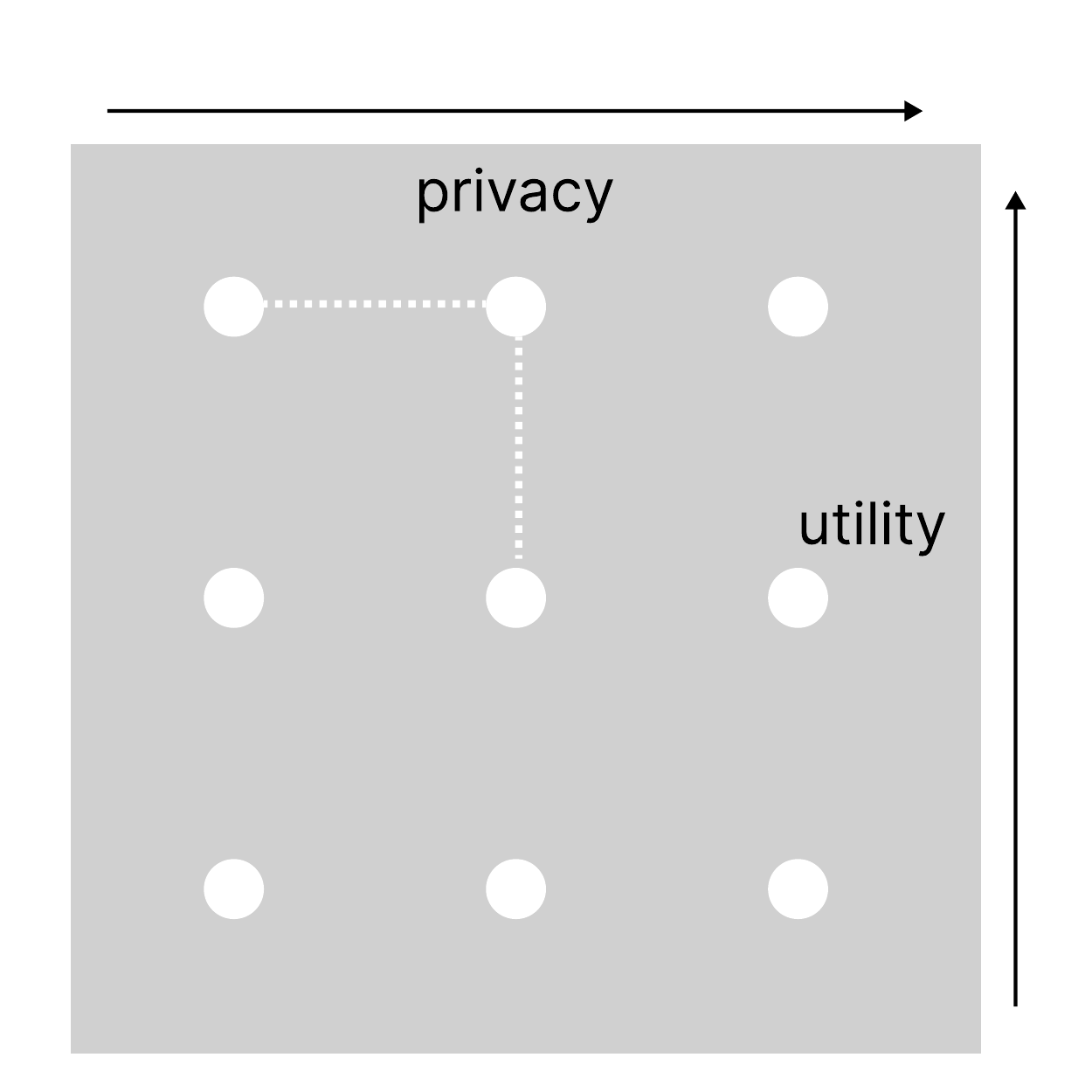}
    }
    \subfloat[Continuous.]{
        \includegraphics[page=3,width=0.3\textwidth]{figures/Adanonymizer.pdf}
    }
    \caption{Different layout designs of the 2D palette.}
    \label{fig:layout}
\end{figure}

For the 2D palette, it has different layout manners and mapping strategies to balance privacy and utility. Regarding the dimension of layouts, the palette could adopt a discrete layout or continuous layout. Discrete layout has the advantages of granting users easier control through choosing from a few options, while continuous layout could guarantee users a more smooth journey through tuning the optimal points, while in the same time observing the change of output in real time, as shown in Figure~\ref{fig:layout}. 

We evaluated several strategies for mapping privacy and utility. One option involved directly mapping the values onto the 2D palette, either linearly or with equal intervals. Linear mapping assigns equal intervals to privacy items with similar sensitivity scores, reflecting users' perceived differences. In contrast, equal interval mapping assigns the same spacing regardless of sensitivity, simplifying item selection. Alternatively, piecewise mapping could be used, where high-privacy and high-utility items are assigned larger intervals than low-privacy and low-utility ones, reflecting participants' priorities. Based on a pilot study, we chose linear mapping, as participants preferred its transparency, which enhanced their sense of control.


\subsection{Interaction Flow and Interface}\label{sec:interface}
The interface of Adanonymizer features a simple and efficient design~\cite{kim2009designing}, functioning as a floating plugin for the input text, making it adaptable to text-based chat and search platforms. Upon selecting the input text, users could activate Adanonymizer, with the selected text as the original input. Adanonymizer consists of three parts (see Figure~\ref{fig:teaser}): a control panel for proactive control through dragging (see Figure~\ref{fig:teaser} (1)), an edit box for showing and editing the output (with changes highlighted, see Figure~\ref{fig:teaser} (2)), and a click button to replace the original input (see Figure~\ref{fig:teaser} (3)). The control panel provides an intuitive way for users to control privacy settings and model output performance levels, similar to a color selection tool. The privacy protection and model output performance levels are marked along the two axes of the panel. Users can adjust the privacy protection and model performances levels through clicking on the 2D palette. Based on these settings, Adanonymizer generates an anonymized version of the text, with variations depending on the level of anonymization users selected, and adopting to different phrases. The edit box features minimal formatting to reduce cognitive load and facilitate seamless system integration, a common design choice in previous literature (e.g., human-AI collaborative writing~\cite{lee2024design,lee2022coauthor}). Users could click on the ``See Label'' button to view the categories of the changed text, which were also highlighted. Users could also directly edit in the output editing box. When users click the ``replace text'' button, the original text is replaced by the anonymized version. 



\begin{figure}
    \centering
    \raisebox{0.5\height}{\subfloat[Multi-constraint sliders.]{
        \includegraphics[width=0.3\textwidth]{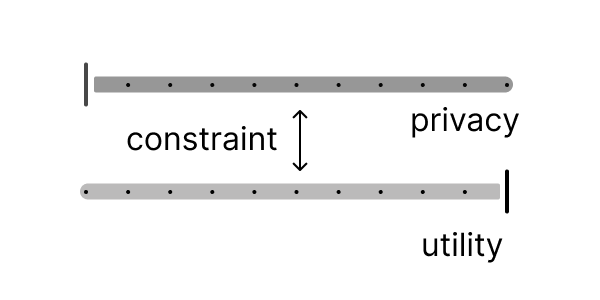}
        \label{fig:single_multi}
    }}
    \raisebox{0.5\height}{\subfloat[Single slider.]{
        \includegraphics[width=0.3\textwidth]{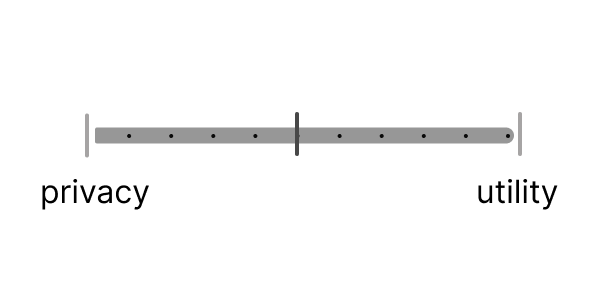}
        \label{fig:single}
    }}
    \subfloat[Panel.]{
        \includegraphics[width=0.3\textwidth]{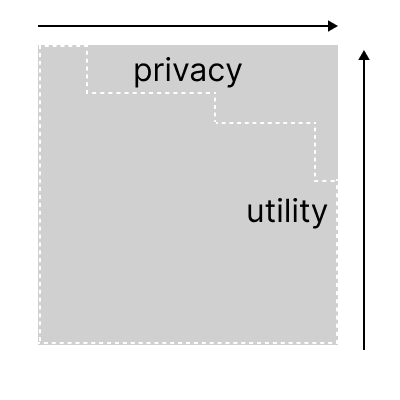}
        \label{fig:panel}
    }
    \caption{The candidate interface designs for privacy-utility balancing.}
    \label{fig:interface}
\end{figure}

Specifically, for the control panel, we sequentially determined the design and the control granularity. We implemented a two-dimensional palette-based design~\cite{shi2022colorcook} for its intuitiveness and simplicity, allowing users to proactively control both privacy protection and model performance. The design was chosen from three options: 1) dragging two progress bars, 2) dragging a single bar with one end prioritizing privacy and the other prioritizing performance, and 3) clicking on a two-dimensional palette to indicate the privacy-performance trade-off. The trade-offs between privacy protection and model performance for three designs are detailed below:

\noindent 1. (Multi-constraint sliders) Dragging the first progress bar restricts the range of the second (Figure~\ref{fig:single_multi}).

\noindent 2. (Single slider) A point on the single bar indicates different trade-offs between privacy and performance (Figure~\ref{fig:single}).

\noindent 3. (Panel) In the two-dimensional palette, the upper-right points are unreachable, showing that privacy protection and model performance cannot be achieved simultaneously (Figure~\ref{fig:panel}).

We selected the third design as it most clearly and intuitively demonstrates the privacy-performance trade-off, allowing users to ``see'' optimal potential cases and more quickly converge to the upper right corner. On contrast, the first design lacked intuitive feedback, requiring users to adjust one bar to see changes in the other. The second design did not clearly and intuitively represent the trade-off, making it difficult for users to understand how privacy protection impacted model performance. In the third design, the horizontal axis represents perceived privacy control (correlated to privacy risk), while the vertical axis reflects perceived model performance. This clear visualization helps users understand the trade-off and make informed decisions. Participants could easily navigate along the trade-off line (as shown by the blue line in Figure~\ref{fig:teaser}) to balance privacy protection and model performance, while benefiting from the intuitive experience of the 2D interface.

For control granularity of the control panel, we opted for continuous rather than discrete control (e.g., splitting into 3$\times$3 chunks) due to the uneven distribution of private information on the palette. Discretizing the palette would have been less effective, as users tend to favor points near the privacy-utility trade-off curve. Continuous control also supports personalized strategies, providing users greater flexibility in balancing privacy and performance. 






\section{Implementation of Adanonymizer}
Adanonymizer is implemented in the form of a local plugin to prevent privacy and ethical issues. We used qwen-2.5-7b-instruct for implementation although the backbone model could potentially be transfered to other architectures. The implementation used Python and Javascript. In this section, we first outline its framework, followed by a detailed description of the anonymization process.


\subsection{System Structure}
The system contained the panel for users to control, the mapping from users' control to the selective anonymization strategy (denoted as \textbf{mapping strategy}) and the back-end anonymization algorithm (denoted as \textbf{anonymization algorithm}). 



\subsection{Mapping Strategy}

The mapping from users' pointing to selective anonymization was computed as described in Section~\ref{sec:study1_tradeoff}. Each point on the curve in Figure~\ref{fig:teaser} corresponds to a specific level of privacy anonymization, showing that higher privacy sensitivity requires stronger anonymization while preserving model performance as much as possible. Let ${X_i}$ denote the information to be anonymized, where $i$ indexes the relevant information. The privacy sensitivity of $X_i$ is represented by $P_i$, and the performance impact (inversely related to the model's reliance on $X_i$) by $M_i$. For a given privacy level $p$, the point on the curve is $(p, \min_{P_i > p} M_i)$. Varying from the minimum to maximum privacy protection levels generates the privacy-utility trade-off curve.

We further normalized the score to $[0,1]$ through first subtracting the minimal value and then divide the range. After normalization, the whole information space would be mapped to the two-dimensional panel and participants could have full control.

\subsection{Anonymization Algorithm}

Researchers proposed different anonymization algorithms, typically involving two steps: 1) identifying the sensitive information \cite{lison2021anonymisation}, 2) removing or substituting that information \cite{majeed2020anonymization,lison2021anonymisation}. As this study does not aim to propose a novel anonymization algorithm, we adopted the most advanced existing methods for each step. For the first step, the core of identifying the sensitive information is named entity recognition, where LLMs could achieve comparable accuracy compared with traditional machine learning and deep learning techniques \cite{wang2023gpt,yohannes2022named}. However, in the context of sensitive information recognition, the study lacked the dataset for training a recognition model due to: 1) the training dataset contained multiple sensitive information of participants, which could hardly be adopted, 2) traditional pre-trained model \cite{hessel2021effective} and softwares (e.g., presidio\footnote{\url{https://www.presidio.com/}}) could not satisfy the recognition task of LLMs input because the input context and PI types changed. Thus, we adopted few-shot prompting on LLMs to identify the sensitive information. For the second step, namely to remove or substitute the sensitive information, existing methods included paraphrase \cite{niu2021unsupervised}, pseudo-anonymize \cite{eder2019identification}, delete the information \cite{lison2021anonymisation} and replacing with the entity types \cite{lison2021anonymisation}. pseudo-anonymization \cite{eder2019identification} was chosen for its lower computational requirements and high inference accuracy. 

The both processes are implemented using a few-shot prompting template. Specifically, based on the results of our pilot study, we opted for a one-shot prompt, as zero-shot prompting did not provide satisfactory accuracy, while using more than two-shot prompts introduced unnecessary latency and cost. The prompt follows structure is: 
\begin{equation}
    P_{\text{guidance}} + E_{\text{one-shot-input}} + E_{\text{one-shot-output}}
\end{equation}
where $P_{\text{guidance}}$ is the guidance and the description of the sensitive information classes. $E_{\text{one-shot-input}}$ is the one-shot prompt input containing various sensitive information. $E_{\text{one-shot-output}}$ is the one-shot prompt output containing identified information and the pseudo-anonymized result. Notably, the one-shot-input and one-shot-output was written and checked by experimenters, thus did not contain real person's information. The detailed implementation was detailed in Section~\ref{app:prompt_design} in the appendix. The model was prompted to output specific entities and their categories such as education and work information, personal basic information, etc. 

To instantiate the recognition process locally, we adopted a qwen-2.5-7b-instruct because of its superior performance on open-source benchmarks. To make the processing possible on personal laptops, we further adopted quantization and runtime optimization techniques, involving using int8 quantization. 

\section{Study 2: Evaluating Adanonymizer in Real-life Tasks}\label{new_study_2}

To explore user perceptions of the anonymization process and facilitate the construction of privacy-utility trade-off relationship, we conducted a user study using real world scenarios to measure perceived privacy risk and utility degradation under different cases.

\subsection{Participants and Apparatus}

We recruited 36 participants (21 males, 15 females, aged 18 to 26, M=21.8, SD=1.6) in China from the XX campus (anonymized for submission) through snowball sampling \cite{goodman1961snowball}. All participants were familiar with interacting with LLMs and had a self-reported LLM usage experience of 8.2 months (SD=2.4). However, they were not familiar with the technical perspective of privacy preservation. No participant dropped the experiment and each participant received 150 RMB as the compensation. The experiment was conducted through an online environment provided as a web service developed by the experimenter through Flask. We implemented the experiment platform using the API of GPT-4o, while we implemented the anonymizers in the form of pop-ups based on the experiment platform.

\subsection{Experiment Design}\label{sec:study2_design}

We conducted a two-factor within-subjects study with the anonymization technique and task as two within factors. Adanonymizer and its ablation techniques operate at three different interactive levels. We set three levels as three different techniques during comparison: \textbf{No interactive (denoted as Automatic)}, \textbf{Single interactive (denoted as Privacy Only)} and \textbf{Full interactive (Adanonymizer)}. In no interactive, the system would modify all the information automatically without user input. In ``Privacy Only'' setting, the user could control the privacy protection level and the system would control which information to modify under the specific privacy protection level. It always maximize utility during modification. In Adanonymizer setting, the user should control both the privacy protection level and model performance preservation level. Notably, the No Interaction and Single Interaction settings serve as ablation comparisons to the Full Interaction setting.

\begin{figure}[!htbp]
    
    \includegraphics[width=0.95\textwidth]{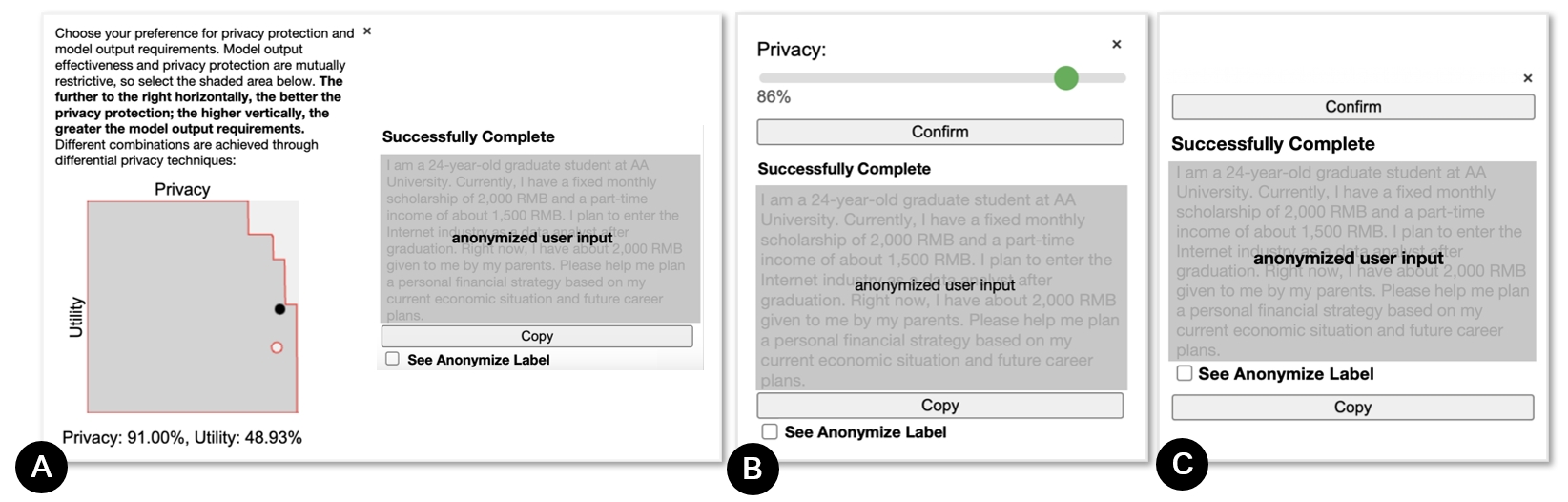}
        
    
    \caption{The experiment interface for (a) Adanonymizer, (b) Privacy Only, (c) Automatic and DP baseline.}
    \label{fig:experiment_platform}
\end{figure}

For the baseline technique, we implemented a state-of-the-art anonymization method using differential privacy, referred to as the \textbf{DP baseline}. The parameters of the differential privacy algorithms were derived from previous research \cite{yue2021differential,chen2022customized}. Thus, four techniques were compared in total: \textbf{Automatic}, \textbf{Privacy Only}, \textbf{Adanonymizer}, \textbf{DP baseline}. For different techniques, we reproduced the original papers' results and implemented them in the form of plug-ins for fair comparison. The detailed interfaces of each technique was shown in Figure~\ref{fig:experiment_platform}. The baseline-DP adopted the state-of-the-art technique from the survey \cite{zhao2022survey}, which implemented a differential privacy based technique with Euclidean distance. We adopted the optimized differential privacy parameters as in the original paper \cite{yue2021differential} and fixed the parameters as users were proved hard to understand the parameter of differential privacy~\cite{nanayakkara2023chances}. We did not implement other baseline techniques due to the following reasons: 1) other anonymization techniques such as training-based methods \cite{hassan2019automatic} and encryption based methods \cite{zhang2024privacyasst} are complicated, which require high computational resource, 2) participants could hardly understand the effectiveness and privacy protection effect as the output were not human-readable text \cite{huang2020texthide}.

We selected the personal consultation tasks to test the anonymization techniques. Personal consultation refers to instances where individuals seek advice or information related to personal matters. Personal consultation could contain varied disciplinary, such as medical consultation, academic documents polishing, writing personal letters or others. Following previous studies \cite{zhang2024s}, we categorized three main scenarios: academic-related, work-related and life-related. Each participant completed all three scenarios using each of the four anonymization techniques, totaling 12 trials (3 scenarios $\times$ 4 techniques). We presented participants the templates and the discussion topics for each scenario. The templates and discussion topics were selected from ShareGPT90K Chinese \footnote{\url{https://huggingface.co/datasets/liyucheng/ShareGPT90K}, last accessed at Jul 30, 2024} and was anonymized by experimenters. We also conducted a pilot study, filtering out those which participants though un-common, hard to understand or hard to think out. Participants in the study needed to craft the input based on the templates and the topics we provided. They needed to write according to their own situation, which ensured they have the precise privacy protection perception and model output perception. However, they were asked to properly replace their own sensitive words with similar but different words (e.g., replace their own name with pseudo names). This reduced the privacy risk during the input. The templates and topics were shown in the appendix in Section~\ref{sec:template}.

Since participants used different LLM products, we standardized the model output performance using a unified experiment platform powered by GPT-4o to facilitate fair comparison (See Figure~\ref{fig:experiment_platform}). We used the API version GPT-4o-2024-05-13. To avoid that the context influence the dialogue and model performance, chat histories and contexts were not maintained. Participants entered their input in the system, and the backend returned the corresponding results.

The experiment proceeded as follows: participants first entered their crafed input, then used the assigned anonymization plug-in to anonymize the text. After entering the anonymized text, they viewed the outputs for both the original and anonymized inputs. Finally, participants modified their input until satisfied and submitted their revised input.

We took the following measurements in the study:

\textbf{Users experience of different techniques}: it is measured through user experience questionnaire - short version (UEQ-S)~\cite{schrepp2017design} and original scale were translated to a Chinese version.

\textbf{Interaction time}: the time is measured from the right clicking until all the user was satisfied with the modification and end the modification process.

\textbf{Modification behavior}: we logged the number of characters edited and tracked user interactions to analyze their modifications following the automatic anonymization.

\textbf{User perception}: we measured five aspects: perceived model output performance, privacy protection effectiveness, satisfaction, perceived control and perceived transparency. All entries are measured using 7-point Likert scale, where 1 indicates the lowest performance, lowest effectiveness, lowest satisfaction, lowest control and lowest transparency and 7 indicates the highest.

\subsection{Procedure}

Participants were first briefed on the experiment and given three minutes to familiarize themselves with the platform. They were then required to sign a consent form before proceeding. The consent form detailed potential risks, including the handling of sensitive inputs and outputs. They were informed that no sensitive data would be collected, and only their interaction behavior with the plug-ins would be analyzed. Participants could withdraw at any time. Each participant completed four sessions, each differing by anonymization technique, with session order counterbalanced using a Latin-square design~\cite{mckay2005number}. In each session, participants interacted with the system for 12 turns (4 anonymization techniques × 3 usage contexts). For each turn, they generated a question related to their personal lives, inputted it, and submitted it to see the anonymized result. Differences were highlighted in yellow, as shown in Figure~\ref{fig:teaser}. Participants could view the categories of private information by clicking the 'See Label' button and compare them. They then submitted the anonymized input to view the output. Participants rested for 90 seconds between sessions to reduce fatigue and maintain study quality~\cite{aaronson1999defining}. After each turn, they rated satisfaction, perceived privacy protection effectiveness, and perceived model output performance. At the end of each session, they completed a questionnaire evaluating the technique, followed by an exit interview after the entire experiment.

\subsection{Results}\label{sec:study2_results}

In this section, we present the results of the different anonymization techniques. We used Repeated Measures Analysis of Variance (RM-ANOVA) for statistical testing and conducted post-hoc comparisons using the Friedman test with Bonferroni adjustment. For non-parametric tests, we applied the Friedman test and used the Nemenyi method for post-hoc comparisons.

\subsubsection{Modification Time}\label{sec:result_modification}

\begin{figure}[!htbp]
    \subfloat[Work.]{
        \includegraphics[width=0.34\textwidth]{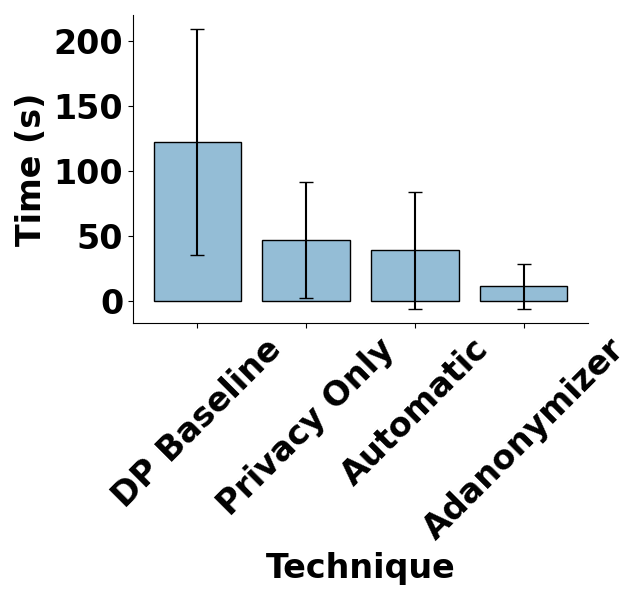}
        \label{fig:study2_work}
    }
    \subfloat[Academic.]{
        \includegraphics[width=0.32\textwidth]{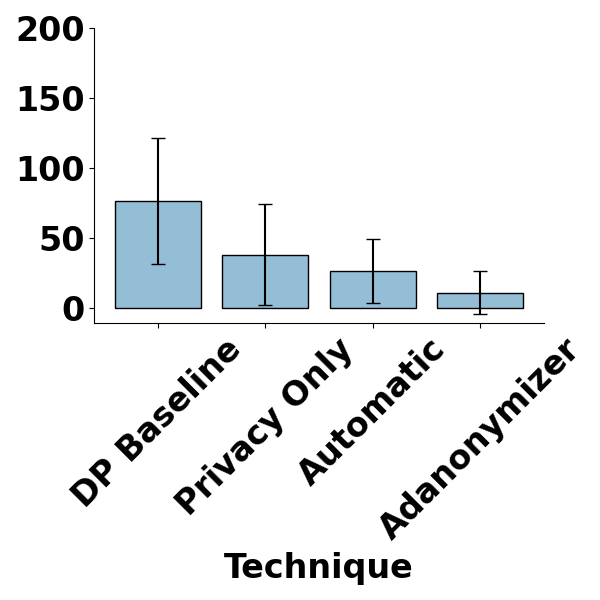}
        \label{fig:study2_academic}
    }
    \subfloat[Daily.]{
        \includegraphics[width=0.32\textwidth]{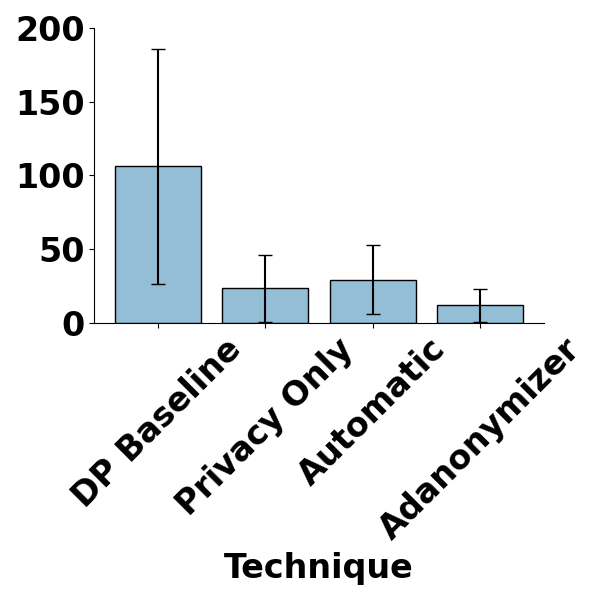}
        \label{fig:study2_daily}
    }
    \caption{The modification time of different techniques for (a) Work, (b) Academic and (c) Daily scenarios. Errorbar indicated one standard deviation.}
    \label{fig:modification_time}
\end{figure}

Modification time was measured from the time users began manual adjustments till they completed their modifications. This metric reflected user satisfaction with the automatic anonymization process-the shorter the modification time, the more effective the automatic process. Specifically, because for all techniques and question length the system latency was shorter than ten seconds and were correlated with the specific questions, we did not count the system latency. Figure~\ref{fig:modification_time} showed the modification times for each technique. We found that Adanonymizer (M=11.5, SD=15.2) significantly reduced the completion time ($F_{3, 105} = 11.5$, $p < .001$) compared to other techniques such as DP baseline (M=101.8, SD=72.8), Privacy Only (M=35.3, SD=35.6) and Automatic (M=31.7, SD=31.9). Post-hoc Tukey HSD comparisons revealed significant differences between different techniques (Automatic v.s. DP Baseline: $p < .001$, DP Baseline v.s. Adanonymizer: $p < .001$, DP Baseline v.s. Privacy Only: $p < .001$), indicating higher user satisfaction with Adanonymizer and less need for manual adjustments.

Across different scenarios, the effect of technique on modification time was also significant (Work: $F_{3, 105} = 6.11$, $p < .001$; Academic: $F_{3, 105} = 6.91$, $p < .001$, Daily: $F_{3, 105} = 9.85$, $p < .001$). In all cases, Adanonymizer consistently outperformed other techniques (all post-hoc $p < .05$), demonstrating its robustness and user preference across varying tasks. 

\subsubsection{Perceived Model Performance and Privacy Protection Effectiveness with Different Scenarios}

\begin{figure}[!htbp]
    \centering
    \subfloat[Work.]{
        \includegraphics[width=0.28\textwidth]{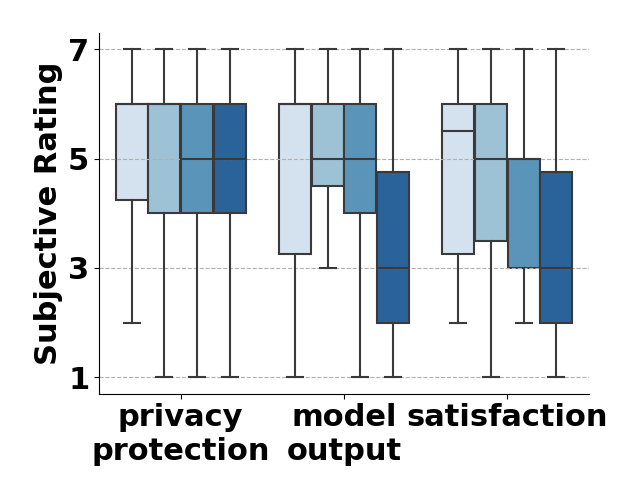}
        \label{fig:study2_subj_work}
    }
    \subfloat[Academic.]{
        \includegraphics[width=0.26\textwidth]{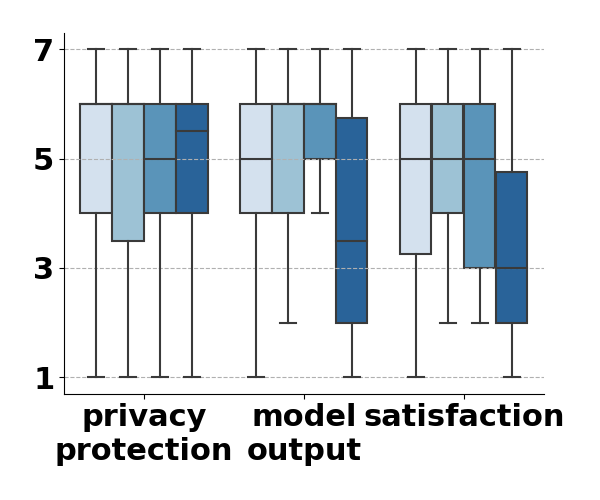}
        \label{fig:study2_subj_academic}
    }
    \subfloat[Daily.]{
        \includegraphics[width=0.44\textwidth]{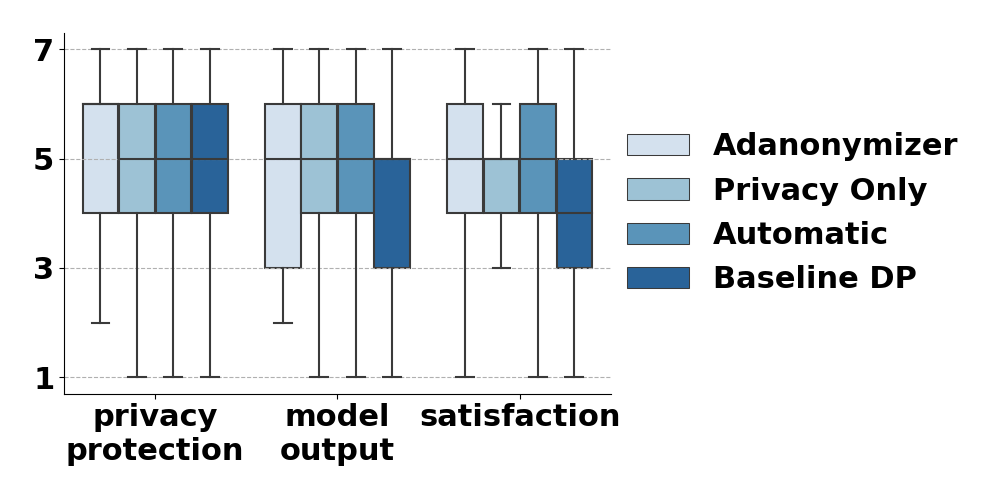}
        \label{fig:study2_subj_daily}
    }
    \caption{The subjective ratings regarding perceived model performance, privacy protection and satisfaction regarding (a) Work, (b) Academic and (c) Daily scenarios (7: most positive, 1: most negative). }
    \label{fig:subj_scenario}
\end{figure}

Figure~\ref{fig:subj_scenario} presents participants' perceived privacy protection across different scenarios. Adanonymizer consistently received higher ratings for both model output and privacy protection compared to other techniques, including the DP-baseline and two ablation methods. Significant effects of techniques on satisfaction were found across all scenarios (Work: $p<.001$, Academic: $p<.001$, Daily: $p<.05$). Post-hoc comparisons showed Adanonymizer significantly outperformed the DP-baseline in all scenarios (Work: $p < .001$, Academic: $p < .01$, Daily: $p < .05$). Adanonymizer also demonstrated superior model output performance in work and academic scenarios (Work: $p < .001$, Academic: $p < .001$), but no significant effect in daily scenarios ($p = .47$). There was no significant differences in privacy protection effect among the techniques (Work: $p = .86$, Academic: $p = .81$, Daily: $p = .27$), indicating that Adanonymizer's performance was comparable with other methods. These findings suggest that while private information in daily scenarios is less frequent and sensitive, Adanonymizer remains robust across contexts.

\subsubsection{Subjective Ratings}
\begin{figure}[!htbp]
    \subfloat[]{
        \includegraphics[width=0.70\textwidth]{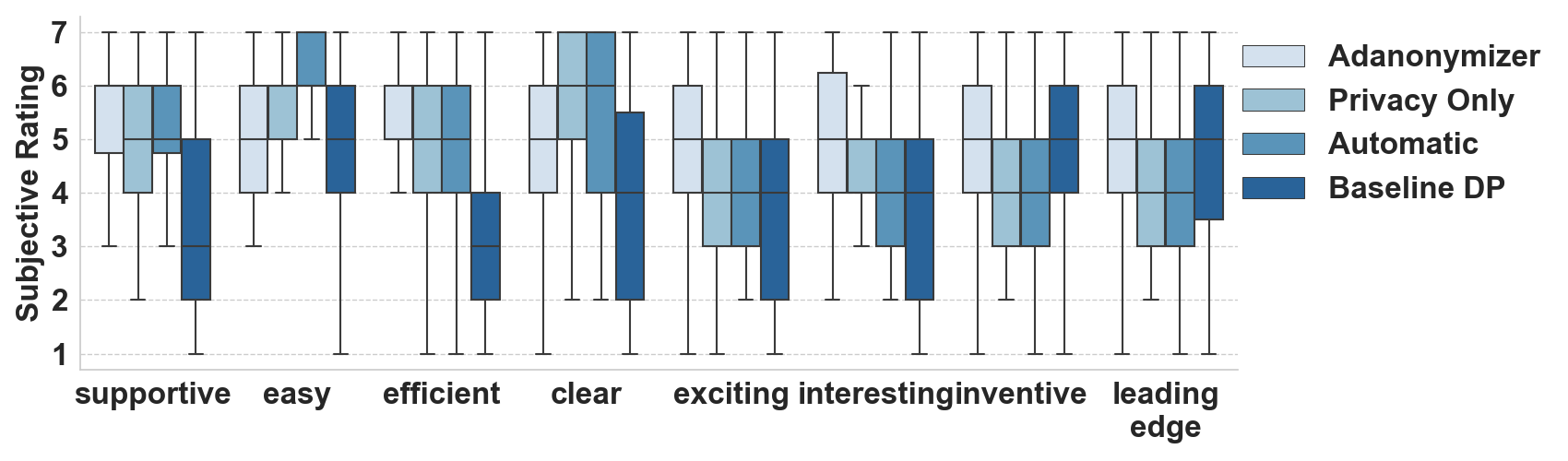}
    }
    
    \subfloat[]{
        \includegraphics[width=0.70\textwidth]{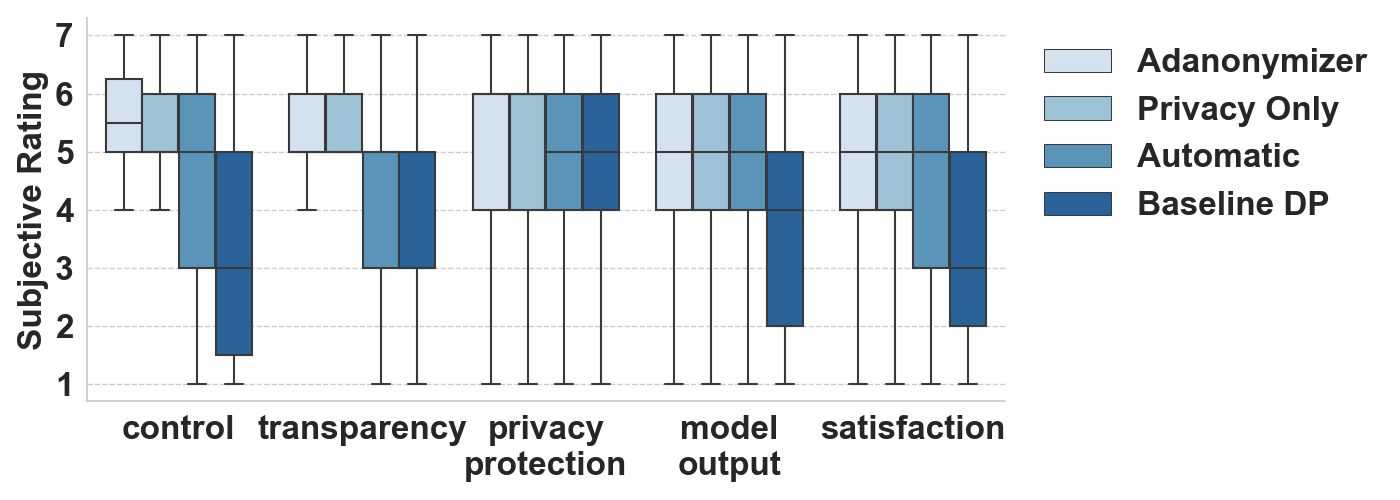}
    }
    \caption{The boxplot ratings for different dimensions, (a) UEQ-S, (b) other subjective ratings.}
    \label{fig:subj}
\end{figure}

Figure~\ref{fig:subj} showed participants' subjective ratings regarding different technique. There are significant differences for all techniques among all dimensions such as Supportive and Easy (Friedman non-parametric test adopted, Supportive: $\chi^2_{3} = 18.3$, $p < .001$, Easy: $\chi^2_{3} = 14.1$, $p < .01$, Efficient: $\chi^2_{3} = 21.1$, $p < .001$, Clear: $\chi^2_{3} = 20.4$, $p < .001$, Exciting: $\chi^2_{3} = 14.6$, $p < .01$, Interesting: $\chi^2_{3} = 8.76$, $p < .05$, Inventive: $\chi^2_{3} = 11.6$, $p < .01$, Leading edge: $\chi^2_{3} = 11.6$, $p < .01$). Post-hoc comparisons found significant differences comparing Adanonymizer with the DP baseline technique in the dimensions of Supportive (post-hoc $p < .01$), Efficient (post-hoc $p < .01$) and Exciting (post-hoc $p < .01$). This proved the efficiency and effectiveness of Adanonymizer. Additionally, significant differences were observed between Adanonymizer and both the ``Privacy Only'' and ``Automatic'' settings in Inventive (all post-hoc $p < .05$) and Leading edge (all post-hoc $p < .05$) dimensions. This demonstrated the effectiveness of Adanonymizer's interactive design.

Besides, we conducted the analysis for the perceived control and helpfulness. Friedman non-parametric test found a significant effect of technique on the perceived control ($\chi^2_{3} = 21.9$, $p < .001$) and perceived transparency of Adanonymizer ($\chi^2_{3} = 24.9$, $p < .001$). Post-hoc comparisons found significant difference of Adanonymizer compared with the DP baseline (perceived control: post-hoc $p < .01$, perceived transparency: post-hoc $p < .01$). This showed Adanonymizer could enable users with better privacy control. Adanonymizer has no significant difference in privacy protection effectiveness compared with other techniques ($\chi^2_{3} = 0.61$, $p = .89$). However, significant effects were observed for model output performance ($\chi^2_{3} = 20.0$, $p < .001$) and satisfaction ($\chi^2_{3} = 23.8$, $p < .001$). Post-hoc nemenyi test found significant differences in satisfaction when comparing Adanonymizer with other techniques (compared with Baseline DP: post-hoc $p < .01$, Privacy Only and Automatic: post-hoc $p < .05$), indicating Adanonymizer effectively meets users' anonymization needs and was preferred overall. 

\subsubsection{Subjective Comments}
Participants gave high praise towards Adanonymizer. \textit{``Adanonymizer clarifies the often opaque process of anonymization by allowing users to manually control the level of anonymity'' (P31)}. This user-centric approach to privacy protection \textit{``enhances user engagement and makes the technology more intriguing'' (P12)}. \textit{``Users can adjust the parameters of Adanonymizer as freely as using a palette'' (P18)}. \textit{``This tool provides me with more opportunities to use anonymization plugins rather than passively accepting the output results'' (P26)}. Participants found Adanonymizer to be \textit{``innovative yet practical'' (P36)}, effectively meeting their needs for anonymization and automatic modifications.

However, some participants suggested that Adanonymizer could \textit{``improve its adjustment strategy based on the results, especially regarding the substitution of different words'' (P35)}. Additionally, participants proposed that the \textit{``fine-grained sliders could be replaced with coarser-grained ones, allowing for simpler control over the anonymization effects'' (P33)}.

\subsubsection{Trade-off Between Privacy Anonymization and Utility}

We interviewed participants to understand their trade-off behavior between privacy anonymization and performance.

\paragraph{Control Path}

We summarized three themes and strategies for users' control. 

\textbf{Tuning along the line.} 15 participants followed the highlighted balance line for tuning their control. ``I would first take notice of the line and then select points near the line to reach the optimal balance point for me.'' (P10). 

\textbf{Target for balance than extreme.} 13 participants first searched and stayed on the upper right part of the panel, which indicates the balance between privacy and utility. They thought ``These points were more suitable because they guaranteed privacy and utility to a great extent in the same time instead of only focusing on one.''

\textbf{Focus on angles which indicate Optimized Points.} 20/36 participants always seek for the corners of the polyline rather than other points. Theoretically the optimal convex line should be consecutive along the privacy-utility panel. However, because the discrete nature of text, anonymization and privacy categories, the curve is a polyline rather than a smooth curve. 

\paragraph{Final Balance State}

We found most of the time all participants would achieve the balance state of privacy and utility. However, nuanced patterns exist and themes were summarized below:

Participants select different balance state for different tasks. For study-related tasks 20/36 participants disclose more information and thus select privacy protection higher, however 10/36 participants cared more about the task performance in work and life-related tasks, thus resulting in more task-oriented privacy protection settings.

Participants select different balance state based on different personalities. A few participants (5/36) cared extremely about their privacy and exhibit extreme patterns for some specific tasks, resulting selecting to anonymize all text. Interestingly, no participant only cared about task performance without considering privacy. All participants' selection were near the line or on the line, indicating that participants always sought for the best balance between privacy and utility.

\section{Discussion}
\subsection{The Trade-off Between Privacy Protection and Model Performance}

The past differential privacy techniques rely on noise addition strategies, which although theoretically sound, may not adequately address the varied sensitivity of tasks or user-specific privacy concerns~\cite{fernandes2019generalised}. In contrast, our user-centric method models privacy preferences based on user ratings, aligning with users preferences better. By examining user perceptions and requirements for 14 detailed data categories and 74 data items, we achieve a finer granularity in balancing the trade-off, offering new insights of users' privacy disclosure perceptions.  The quantification of this trade-off not only reveals the inherent flexibility in privacy management but also highlights important findings. Specifically, the results suggest that privacy protection, when designed with user control in mind, does not always result in performance degradation, which provides the opportunity to optimize both privacy protection and model performance.

\subsection{Feasibility of Adanonymizer}

Adanonymizer employed a proactive design, allowing users to control both privacy protection and model performance levels. This approach enhanced user agency \cite{lukoff2021design,coyle2012did} and satisfaction \cite{zhang2019proactive}. By involving users in the loop, the design addressed the limitations of objective privacy metrics (e.g., differential privacy) that often fail to capture users' perception of privacy protection and model performance \cite{staab2024large}. During experiment, users found the automatic mode too restrictive, as it did not allow easy adjustment of privacy levels. Moreover, they were often uncertain about the model’s performance, leading to reluctance in modifying privacy settings. In contrast, Adanonymizer, the manual mode, provided intuitive control over both aspects through a simple drag-and-drop interface, which users appreciated for its simplicity and ease of use.

We tested the effectiveness of Adanonymizer using personal consultation tasks (see Section~\ref{sec:study2_design}) and found frequent private information leakage. As the personal private information also appears in other tasks such as collaborative writing~\cite{lee2024design}, and text editing~\cite{lee2022coauthor}, we envisioned the anonymization effectiveness and the model output performance maintenance of these tasks could also satisfy users' need. Although Adanonymizer was designed to be used in LLMs' input, the form of plug-in and the anonymization effect enable it to be easily transformed to other scenarios, such as text document anonymization \cite{lison2021anonymisation}, anonymization during searching \cite{peddinti2011effectiveness} or anonymization during medical consultation \cite{abbott2019local}. While the current implementation utilizes prompt-based methods for large models, which achieve comparable results to NLP techniques, it is also feasible to train offline models to attain similar outcomes when synthetic data or compliant original data is available.

\subsection{Design Implications}

This paper leveraged the collaborative metaphor to manage the privacy information during human-LLM interaction. Based on these, we provided several key implications: 

\textbf{Enabling users to proactively control the privacy-utility balancing based on privacy risk and utility modeling.} Based on privacy calculus theory~\cite{laufer1977privacy}, application fields such as healthcare~\cite{ragan2018balancing} could leverage the modeling framework to engage users in proactively control their trade-off between privacy and utility, thus enhancing the automatic algorithmic-based balancing methods.

\textbf{Interaction designs could leverage interactive 2D panels for balancing trade-offs such as privacy protection and model performance.} 

Future interaction design for multimodal frameworks~\cite{hu2024exploring} or privacy protection for visual language models~\cite{wang2023modeling} could leverage similar framework to manage the privacy-utility trade-off through the 2D panels, enhancing transparency and ease-of-control.


\textbf{Privacy protection methods can enhance their transparency through controllable panel metaphors.}
Many algorithms, such as DP-based methods~\cite{xiong2020towards,franzen2022private}, are inherently opaque and require clear communication of their parameters, settings, and effects to users. The panel based approach to visualize privacy protection and model performance can also be extended to other methods, such as DP~\cite{fernandes2019generalised} or encryption~\cite{lin2024promptcrypt}, by illustrating $\epsilon$ in DP and cryptographic strength in encryption algorithms, where user control could determine the privacy budget and govern the background protection algorithm.

\section{Ethical Considerations}
We acknowledged that our research may have ethical issues. We followed Menlo report \cite{bailey2012menlo} and Belmont report \cite{beauchamp2008belmont} in designing the studies and tried our best to avoid ethical concerns. In all studies, we compensated participants according to the local wage standard and told the participants at the beginning of the experiment about the potential benefits and harms. Participants were allowed to quit at any time in the experiment if they felt uncomfortable or for other reasons. Our experiment aimed at solving the privacy policy reading problems through designing applications to facilitate reading. The participants may potentially benefit from reading the privacy policy as they could acquire more information regarding their personal information collection. Besides, all the participants' experimental data was stored on a local device with encryption.

\section{Limitation and Future Works}
We acknowledged that our study have limitations and regard these as future directions. First, although we tried to diversify the background of the participants, we recruited the participants through snowball sampling in the campus, which restricted the age and educational background. University students was a group with relatively higher educational background than the average \cite{betts1999determinants} and may understand privacy policy easier. This would better prove the problems during the reading of privacy policy and the effectiveness of our Adanonymizer. 

Second, the experiment may face social desirability \cite{chung2003exploring} and recall bias \cite{coughlin1990recall} of the participants, during which participants may say more positive opinions towards privacy policy and the reading issues. This proved that in reality, the problems regarding reading privacy policy may be greater. 

Third, we only selected the privacy risk towards re-training \cite{zhang2024s} to study. The most significant difference of human-LLM interaction the traditional online media interaction \cite{bryce2014role} and human-AI interaction \cite{amershi2019guidelines} was that LLMs frequently needed novel data for fine-tuning and model improvement. Although there exist other data misuse and privacy risk beyond re-training, different usage exhibit different privacy risk patterns for participants, which were out of this paper's scope. We deemed those as our future work. Moreover, we utilized large models for Adanonymizer to avoid using users' private data. In future deployments, companies could train models with compliant data to achieve the same functionality.

\section{Conclusion}

Users suffered from input a lot of private information during the interaction of LLMs. Previous studies hoped to achieve both model performance and privacy protection, however failed due to a lack of human-centric modeling. To solve this problem, we first conducted a vignette study asking users to rate the privacy risk and performance influence of different personal information. We then proposed Adanonymizer, which utilized the users' perception of privacy risk and utility degradation. Adanonymizer was designed in a manner that users could proactively control the privacy protection level through dragging bar and implemented through prompt-based methods. In the evaluation study, Adanonymizer out-performed other baseline techniques in terms of privacy protection, model performance and completion speed. It is also preferred for its proactive control and ease of understanding.

\bibliographystyle{ACM-Reference-Format}
\bibliography{sample-base}

\appendix 
\section{Prompting Implementation of Anonymizer}\label{app:prompt_design}
The prompt consisted of three parts. The first part described the rules including the description of information, the formatting, etc. The second part is a one-shot prompt input containing various forms of sensitive information. The third part is the anonymized output of the one-shot prompt input. The content of the three parts is as follows:

\subsection{Rules and Guidance}
Please act as an expert and analyze the private information in the below paragraph. I'll give you an example first.

[Information Classes]

Name: the name of some person.

Birthday: the birthday of some person.

Age: number of years lived.

Gender: male or female or other.

Ethnicity / Race: ethnic background. 

Nationality: country of citizenship.

Place of origin: hometown or birthplace.

Marital status: single, married, etc.

Family relationships: relatives and their connections.

Address: residential location.

Phone number: contact telephone number.

Email address: electronic mail address.

Hobbies and interests: personal interests and activities.

Identification card: national ID number.

Passport: passport number.

Driver's license: license to drive.

Work permit: employment authorization.

User account: account username.

User ID: unique user identifier.

Instant messaging account: IM service handle.

Social media account: social network handle.

Nickname: informal name or alias.

IP address: internet protocol address.

Weight: body mass.

Height: stature.

Blood type: blood group classification.

Medical conditions: health status and issues.

Medical instructions: doctor's orders.

Test reports: medical test results.

Physical examination reports: health checkup results.

Medical history: past health issues.

Educational background: level of education completes.

Degree: academic degree.

Educational experience: schooling history.

Transcript: academic record.

Occupation: job title.

Job title: specific position at work.

Employer: place of employment.

Work location: place of work.

Work experience: employment history.

Salary: earnings from employment.

Resume: professional CV.

Bank card number: debit/credit card number.

Payment account: online payment account.

Account balance: available funds.

Transaction order: purchase or sale record.

Transaction amount: value of transactions.

Payment records: history of payments.

Income status: financial earnings.

Property information: real estate details.

Deposit information: savings details.

Vehicle information: car ownership details.

Tax amount: taxes paid.

Virtual property: digital assets.

Loan information: borrowing details.

Repayment information: loan repayment status.

Debt information: outstanding debts.

Credit records: credit history.

Credit information: credit score and details.

Account login password: password for account access.

Bank card password: PIN for bank card.

Payment password: password for transactions.

Account query password: password for account 
Transaction password: password for confirming transactions.

Bank card verification code: CVV for security code.
USB key: security token device.

Dynamic password: one-time password.

SMS verification code: code sent via SMS for verification.

Personal digital certificate: digital ID for security.

Random token: security token for access.

Communication records: logs of communications.

SMS: text messages.

Email: electronic mail messages.

Instant messaging: online chat records.

Contacts: address book entries.

Friends list: list of social connections.

Group list: list of group memberships.

Email address list: collection of email contacts.

Family relationships: relatives and their connections.

Work Relationships: colleague connections.

Social Relationships: friends and acquaintances.

Web Browsing History: record of internet sites visited.

Software Usage Records: application usage data.

Cookies: web browsing data files.

Published Social Information: posted social media content.

Search History: record of internet searches.

Download History: record of downloaded files.

Region Code: geographical area code.

City Code: code for city identification.

Longitude and Latitude: geographic coordinates.

Accommodation Information: housing details.

Community Code: neighborhood identifier.

Step Count: number of steps taken.

Step Frequency: steps per minute.

Exercise Duration: length of workout.

Exercise Distance: distance covered in exercise.

Exercise Type: type of physical activity.

Heart Rate during Exercise: beats per minute during exercise.

Sexual Orientation: sexual preference.

Marriage History: past marital status.

Religious Belief: faith or religion.

Undisclosed Criminal Records: hidden criminal history.

Common Languages: languages spoken.

Past or Current Educational Majors: field of study.

\subsection{One-shot Prompt Input Example}
[Example input]
John Doe, born on January 1, 1980, is a 44-year-old Caucasian male from New York, USA. He holds American nationality and currently resides at 1234 Elm Street, Springfield, IL, 62704. John is married to Jane Doe, and they have two children, Alice and Bob Doe. He enjoys reading, hiking, and photography in his free time.

John's contact information includes a phone number, (123) 456-7890, and an email address, john.doe@example.com. His identification details include a national ID (123456789), a passport (A12345678), a driver's license (D1234567), and a work permit (WP123456). For online activities, he uses the user account johndoe\_80 with user ID 001234567, and his instant messaging and social media accounts are johndoeIM and \@johndoe, respectively. He is often referred to by his nickname, Johnny, and his IP address is 192.168.1.1.

Physically, John is 6 feet tall, weighs 180 pounds, and has an O+ blood type. He has a medical history of asthma and currently manages hypertension with daily medication. His latest medical reports, including blood test and physical examination, indicate good health.

John holds a Bachelor's degree in Computer Science from the University of Illinois, where he studied from 1998 to 2002. Professionally, he is a Senior Developer at Tech Solutions Inc., located at 5678 Oak Street, Springfield, IL, 62704, and has 20 years of experience in software development, earning an annual salary of \$120,000. His resume and academic transcript are available upon request.

Financially, John manages a bank card (1234 5678 9012 3456) and an online payment account (johndoe\_pay@example.com) with a current balance of \$5,000. He has conducted transactions such as a \$200 order and a \$50 payment to Amazon. He owns a house at 1234 Elm Street, a 2018 Toyota Camry, and has a savings account with \$10,000. His annual tax payment is \$12,000, and he holds virtual assets, including 5 Bitcoin. John has a mortgage of \$200,000, with a monthly repayment of \$1,500, and a credit card debt of \$2,000, but maintains an excellent credit score of 750.

For security, John uses various passwords, a USB key, dynamic passwords, SMS verification codes, a personal digital certificate, and a random token for account access and transactions. His communication records, including SMS, email, instant messaging, and contacts, are archived. He has 150 friends on social media, is a member of 10 online groups, and maintains an email address list for personal and professional use.

John's browsing history, software usage records, cookies, published social information, search history, and download history are recorded. His geographic information includes a region code (123), city code (456), and coordinates (39.7817° N, 89.6501° W). He has accommodation details for travel, a community code (789), and tracks his physical activity, averaging 10,000 steps daily at a frequency of 100 steps per minute, with 30 minutes of exercise covering 3 miles. His heart rate during exercise is typically 140 BPM.

John identifies as heterosexual, has been married since 2005, and practices Christianity. He has no undisclosed criminal records and is fluent in English and Spanish. His academic focus has always been in Computer Science.

\subsection{One-shot Prompt Output Example}
(John Doe)[Name], born on (January 1, 1980)[Date of Birth], is a (44-year-old)[Age] (Caucasian)[Ethnicity] (male)[Gender] from (New York, USA)[Place of Origin]. He holds (American nationality)[Nationality] and currently resides at (1234 Elm Street, Springfield, IL, 62704)[Address]. John is (married)[Marital Status] to Jane Doe, and they have two children, (Alice and Bob Doe)[Family Relationships]. He enjoys (reading, hiking, and photography)[Hobbies and Interests] in his free time.

John's contact information includes a phone number, ((123) 456-7890)[Phone Number], and an email address, (john.doe@example.com)[Email Address]. His identification details include a national ID ((123456789))[Identification Card], a passport ((A12345678))[Passport], a driver's license ((D1234567))[Driver's License], and a work permit ((WP123456))[Work Permit]. For online activities, he uses the user account (johndoe\_80)[User Account] with user ID (001234567)[User ID], and his instant messaging and social media accounts are (johndoeIM)[Instant Messaging Account] and (@johndoe)[Social Media Account], respectively. He is often referred to by his nickname, (Johnny)[Nickname], and his IP address is (192.168.1.1)[IP Address].

Physically, John is (6 feet tall)[Height], weighs (180 pounds)[Weight], and has an (O+ blood type)[Blood Type]. He has a medical history of (asthma)[Medical History] and currently manages (hypertension)[Medical Conditions] with daily medication (Medical Instructions). His latest medical reports, including (blood test)[Test Reports] and (physical examination)[Physical Examination Reports], indicate good health.

John holds a (Bachelor's degree)[Degree] in (Computer Science)[Past or Current Educational Majors] from (University of Illinois)[Educational Experience], where he studied from 1998 to 2002. Professionally, he is a (Senior Developer)[Job Title] at (Tech Solutions Inc.)[Employer], located at (5678 Oak Street, Springfield, IL, 62704)[Work Location], and has (20 years)[Work Experience] of experience in software development, earning an annual salary of (\$120,000)[Salary]. His (resume)[Resume] and academic (transcript)[Transcript] are available upon request.

Financially, John manages a bank card ((1234 5678 9012 3456))[Bank Card Number] and an online payment account ((johndoe\_pay@example.com))[Payment Account] with a current balance of (\$5,000)[Account Balance]. He has conducted transactions such as a (\$200 order)[Transaction Order] and a (\$50 payment to Amazon)[Payment Records]. He owns a house at (1234 Elm Street)[Property Information], a (2018 Toyota Camry)[Vehicle Information], and has a savings account with (\$10,000)[Deposit Information]. His annual tax payment is (\$12,000)[Tax Amount], and he holds virtual assets, including (5 Bitcoin)[Virtual Property]. John has a mortgage of (\$200,000)[Loan Information], with a monthly repayment of (\$1,500)[Repayment Information], and a credit card debt of (\$2,000)[Debt Information], but maintains an excellent credit score of (750)[Credit Information].

For security, John uses various passwords ([Account Login Password], [Bank Card Password], [Payment Password], [Account Query Password], [Transaction Password]), a (USB key)[USB Key], dynamic passwords ([Dynamic Password]), SMS verification codes ([SMS Verification Code]), a (personal digital certificate)[Personal Digital Certificate], and a (random token)[Random Token] for account access and transactions. His (communication records)[Communication Records], including (SMS)[SMS], (email)[Email], (instant messaging)[Instant Messaging], and (contacts)[Contacts], are archived. He has (150 friends)[Friends List] on social media, is a member of (10 online groups)[Group List], and maintains an (email address list)[Email Address List] for personal and professional use.

John's (browsing history)[Web Browsing History], (software usage records)[Software Usage Records], (cookies)[Cookies], (published social information)[Published Social Information], (search history)[Search History], and (download history)[Download History] are recorded. His geographic information includes a (region code (123))[Region Code], (city code (456))[City Code], and (coordinates (39.7817° N, 89.6501° W))[Longitude and Latitude]. He has (accommodation details)[Accommodation Information] for travel, a (community code (789))[Community Code], and tracks his physical activity, averaging (10,000 steps)[Step Count] daily at a frequency of (100 steps per minute)[Step Frequency], with (30 minutes of exercise covering 3 miles)[Exercise Distance]. His heart rate during exercise is typically (140 BPM)[Heart Rate during Exercise].

John identifies as (heterosexual)[Sexual Orientation], has been (married since 2005)[Marriage History], and practices (Christianity)[Religious Belief]. He has no (undisclosed criminal records)[Undisclosed Criminal Records] and is fluent in (English and Spanish)[Common Languages]. His academic focus has always been in (Computer Science)[Past or Current Educational Majors].

\section{The Content of the Questionnaire in Study 1}\label{sec:appendix_questionnaire}
The questionnaire contained multiple aspects, which is the privacy risk description, the information content description, the test questions, the questionnaire rating entries and the questions asking participants to disclose more information.

\subsection{Privacy Risk Description}
As the capabilities of LLMs continue to improve, these models can now handle a wider variety of tasks in everyday life, leading to the inclusion of increasing amounts of personal information in the input prompts. If you input specific categories of information (i.e., information of that category appears in text form within the prompt) into a LLM, the following different risk categories may arise.

\textbf{Risk of Information Leakage:} There is a risk of sensitive information being disclosed to malicious third parties by service providers when users share sensitive data such as business information, academic assignments, financial data, legal cases, medical information, and personal life details. For example, if you use downstream products of a large model to polish your resume, the company might leak it to recruitment website like LinkedIn. If you use downstream products of a large model to select appliances, the company might leak your preferences to shopping websites.

\textbf{Risk of Retraining:} Users' sensitive information might be used by service providers to improve models through training. Model training (pre-training or fine-tuning) involves using user data to teach the model how to respond to user queries. In cases where user A's data is used for training, there is a privacy risk if the model outputs user A's data when responding to unrelated or malicious inpus from user B. For instance, if you ask a large model to polish your article draft, the model might leak your draft to user B when generating content.

\subsection{Information Content Description}
The following are the privacy information categories involved in this questionnaire. Below are specific examples illustrating how such privacy information might be input into a LLM in natural language. Note that these examples are provided solely to help you understanding how personal information might be input into the model and do not represent all possible input scenarios. You are free to imagine other potential situations:

\textbf{Job/Study Scenarios}: ``I am a graduate student at Western University, majoring in Industrial Engineering. I am currently applying for summer internships. Here is my resume, and my job preferences include positions such as Algorithm Engineer and AI Product Manager. My main areas of expertise are commercialization and user products. Can you help me polish my resume and write a cover letter? Additionally, please recommend some companies and positions that suit me, and analyze my chances of securing a permanent position based on the company's business development.''

\textbf{Personal Basic Information}: name, date of birth, age, gender, etc.

\textbf{Personal Education Information}: educational background, transcripts, etc.

\textbf{Personal Work Information}: occupation, job title, workplace, etc.

\textbf{Personal Communication Information}: communication records, text messages, etc.

\textbf{Personal Identification Information}: ID card, passport, driver's license, work ID, etc.

\textbf{Social Relationship Scenarios}: ``My father is an associate professor in the Department of History at Tsinghua University. He thinks I can stay in the department after completing my Ph.D., but I found that his department hasn't hired any new faculty in the past five years. Please analyze my chances of securing a faculty position in five years, considering the overall development of the history discipline and departmental relationships.''

\textbf{Personal Social Information}: Contact list, social relationships, etc.

\textbf{Daily Help Scenarios}: ``Hello, based on my recent internet activity, analyze my personality type and provide it in the form of MBTI.''

\textbf{Personal Internet Activity Information}: Web browsing history, software usage records, etc.

\textbf{Personal Preference Information}: Interests and hobbies, favorite music/food/furniture, etc.

\textbf{Health Scenarios}: ``Hello, here is my recent medical report (attached). I have been diagnosed with fatty liver disease, and there is history of diabetes in my family, including my mother, grandmother, and grandfather. I need to lose weight and prevent the risk of diabetes. My current exercise plan is running for 30 minutes twice a week. Please develop a better exercise plan for me. 



\section{The Demographics of Participants in Study 1}\label{sec:study1_demo}

The demographics of participants in Study 1 is shown in Table~\ref{tbl:study1_demographics}.

\begin{table}[!htbp]
\centering
\caption{Demographics of participants in Study 1.}
\label{tbl:study1_demographics}
\begin{tabular}{lrr}
\toprule
\textbf{Demographic Information} & \textbf{Count} & \textbf{Percentage} \\
\midrule
\textbf{Gender} & & \\
Male & 98 & 44.34\% \\
Female & 123 & 55.66\% \\
\midrule
\textbf{Age Group} & & \\
18-25 & 22 & 9.95\% \\
26-35 & 141 & 63.80\% \\
36-45 & 47 & 21.27\% \\
46-55 & 10 & 4.52\% \\
56-65 & 1 & 0.45\% \\
\midrule
\textbf{Occupation} & & \\
Manufacturing (factory workers, mechanical engineers, electrical engineers,\\ 
chemical engineers, etc.) & 56 & 25.34\% \\
Construction (construction workers, architects, civil engineers, interior designers, etc.) & 6 & 2.71\% \\
Information Technology (software development engineers, system administrators, \\ 
network engineers, data analysts, etc.) & 45 & 20.36\% \\
Finance (bank clerks, investment advisors, accountants, financial analysts, etc.) & 21 & 9.50\% \\
Education and Research (teachers, professors, researchers, laboratory technicians, etc.) & 22 & 9.95\% \\
Healthcare (doctors, nurses, pharmacists, medical technicians, etc.) & 7 & 3.30\% \\
Service Industry (waitstaff, hotel managers, customer service representatives, \\ 
beauticians, etc.) & 16 & 7.24\% \\
Retail (salespersons, cashiers, store managers, warehouse managers, etc.) & 9 & 4.07\% \\
Transportation (drivers, pilots, traffic controllers, logistics managers, etc.) & 5 & 2.26\% \\
Public Service (police officers, firefighters, civil servants, social workers, etc.) & 4 & 1.81\% \\
Culture and Entertainment (actors, musicians, writers, directors, etc.) & 1 & 0.45\% \\
Journalism and Media (journalists, editors, photographers, broadcasters, etc.) & 4 & 1.81\% \\
Law (lawyers, judges, legal advisors, notaries, etc.) & 1 & 0.45\% \\
Human Resources (HR managers, recruitment specialists, trainers, compensation \\ 
and benefits specialists, etc.) & 21 & 9.50\% \\
Others & 3 & 1.36\% \\
\midrule
\textbf{Educational Background} & & \\
High school or below & 4 & 1.81\% \\
Associate degree & 14 & 6.33\% \\
Bachelor's degree & 171 & 77.38\% \\
Master's degree or above & 32 & 14.48\% \\
\bottomrule
\end{tabular}
\end{table}

\section{The Templates and Questions used in Study 2}\label{sec:template}

We used the following questions and templates:

\subsection{Work-related Scenarios}

\textbf{Task:} You need the system to help you with problems in your work. Your descriptions of the question and the references should be as detailed as possible. Here are some example questions for inspiration: 

\begin{itemize}
    \item Create a brief cover letter summary for me.
    \item You are seeking GPT's advice on career/internship/future career planning. We hope you find a familiar, daily issue to consult GPT about.
    \item When applying for a job, you need to give GPT your resume and ask for feedback.
    \item You need to prepare for an interview for a social work position and would like GPT to polish your self-introduction speech.
\end{itemize}







\subsection{Academic-related Scenarios}

\textbf{Task:} You need the system to help you with academic related problems. Your descriptions of the question and the references should be as detailed as possible. Here are some example questions for inspiration: 

\begin{itemize}
    \item You need to write an email to a professor with whom you hope to collaborate, and you provided GPT with the following information. 
    \item You need to prepare for studying abroad and want GPT to help you revise your personal statement and resume.
    \item You encounter a bottleneck in your research and hope GPT can help you brainstorm new research topics.
    \item You are preparing for an academic poster presentation and want GPT to help you refine and enhance the content outline of your poster.
    \item You need to write an application form for an outstanding graduate student party member, and you provided GPT with the following information. 
\end{itemize}

\subsection{Life-related Scenarios}




\textbf{Task:} You need GPT to help you with simple daily decisions and hope your descriptions and references are as detailed as possible. Here are some example questions for inspiration:
\begin{itemize}
    \item You need GPT to help you write a year-end summary, weekly report, or monthly report.
    \item You need GPT's advice on social issues, such as resolving conflicts with friends, family, mentors, or supervisors.
    \item Ask GPT to analyze emotional situations or personal relationships.
    \item Based on a recent medical report or your daily routine, create an exercise and health plan.
    \item Based on your family medical history, identify potential health risks and suggest preventive measures.
    \item Analyze potential health and time management issues based on app usage and screen time.
    \item Create an exercise plan based on your recent physical condition and schedule.
    \item Based on the location, years of work experience, and housing prices in a few cities, provide a reasonable mortgage plan or future financial plan.
    \item You want GPT to help you analyze which among a few funds/new properties/stocks has the most growth potential.
    \item Based on your current income, academic status, future career plans, and development path, have GPT help you plan your personal financial route.
\end{itemize}

\section{The Detailed Information Types}\label{appen:information_type}
The mapping of detailed information types and the categories were as follows:

\begin{longtable}{p{6.2cm}p{3.3cm}p{4cm}}
\caption{The detailed information types used in Study 1. Info in subcategories denoted information and Comm. in subcategories denoted Communication} \\
\hline
\textbf{Category} & \textbf{Subcategory} & \textbf{Type} \\
\hline
\endfirsthead
\hline
\textbf{Category} & \textbf{Subcategory} & \textbf{Type} \\
\hline
\endhead
\hline \multicolumn{3}{|r|}{{Continued on next page}} \\ \hline
\endfoot
\hline
\endlastfoot

\multirow{3}{*}{Personal Basic Information} & Basic Info & Name, Date of Birth, Age, Gender, Ethnicity, Nationality,  Place of Origin, Marital Status, Family Relationships, Address, Phone Number, Email Address, Hobbies \\
\hline
\multirow{2}{*}{Personal Identity Information} & Identity Info & ID Card, Passport, Driver’s License, Work ID \\
\hline
\multirow{2}{*}{Online Identity Information} & Online ID & User Account, User ID, Instant Messaging Account,  Social Media Account, Nickname, IP Address \\
\hline
\multirow{2}{*}{Personal Health Information} & Health Info & Weight, Height, Blood Type \\
& Medical Info & Diagnosis, Prescription, Lab Report, Health Report, Medical History \\
\hline
\multirow{2}{*}{Personal Education and Employment Information} & Education Info & Education Level, Degree, Education History, Transcript \\
& Employment Info & Occupation, Job Title, Employer, Work Location, Work Experience,  Salary, Resume \\
\hline
\multirow{3}{*}{Personal Property Information} & Financial Info & Bank Card Number, Payment Account, Account Balance \\
& Transaction Info & Transaction Order, Transaction Amount, Payment Record \\
& Asset Info & Income Status, Real Estate Information, Savings Information,  Vehicle Information, Tax Amount, Virtual Property \\
& Loan Info & Loan Information, Repayment Information \\
\hline
\multirow{2}{*}{Identity Verification Information} & Verification Info & Account Login Password, Bank Card Password, Payment Password,  Account Query Password, Transaction Password, Bank Card Verification Code,  USB Key, Dynamic Password, SMS Verification Code, Personal Digital Certificate, Random Token \\
\hline
\multirow{2}{*}{Personal Communication Information} & Comm. Info & Communication Records, SMS, Email, Instant Messaging \\
\hline
\multirow{2}{*}{Contact Information} & Contact Info & Address Book, Friends List, Group List, Email Address List, Family Relationships, Work Relationships, Social Relationships \\
\hline
\multirow{2}{*}{Personal Internet Records} & Internet Records & Web Browsing Records, Software Usage Records, Cookies, Social Media Posts, Search History, Download History \\
\hline
\multirow{2}{*}{Personal Location Information} & Location Info & Region Code, City Code, Latitude and Longitude, Accommodation Information, Neighborhood Code \\
\hline
\multirow{2}{*}{Personal Activity Information} & Activity Info & Steps, Step Frequency, Activity Duration, Activity Distance,  Activity Mode, Heart Rate \\
\hline
\multirow{2}{*}{Other Personal Information} & Other Info & Sexual Orientation, Marital History, Religious Beliefs, Unpublicized Criminal Record \\
\hline

\end{longtable}

\end{document}